\pdfoutput=1
\documentclass[acmsmall]{acmart}

\newcommand{\synote}[1]{\textcolor{blue}{#1}}

\newcommand{\comm}[1]{\textcolor{gray}{#1}}

\newcommand{\syrevise}[1]{\textcolor{black}{#1}}

\newcommand{\syrevisetwo}[1]{\textcolor{black}{#1}}

\usepackage{graphicx}
\makeatletter
\newcommand*\bigcdot{\mathpalette\bigcdot@{.5}}
\newcommand*\bigcdot@[2]{\mathbin{\vcenter{\hbox{\scalebox{#2}{$\m@th#1\bullet$}}}}}
\makeatother

\usepackage{amsmath, bm}

\usepackage{xcolor}
\definecolor{deleteColor}{RGB}{252,217,217} % 自定义颜色
\definecolor{addColor}{RGB}{217,255,218} % 自定义颜色

\usepackage{colortbl}
\usepackage{multirow}

\usepackage{tcolorbox}
\usepackage{threeparttable}

\usepackage{hyperref}

\usepackage{amsmath}
\DeclareMathOperator*{\concat}{\oplus}

\usepackage{listings}
\AtBeginDocument{%
  \providecommand\BibTeX{{%
    \normalfont B\kern-0.5em{\scshape i\kern-0.25em b}\kern-0.8em\TeX}}}

\setcopyright{acmcopyright}
\copyrightyear{2018}
\acmYear{2018}
\acmDOI{XXXXXXX.XXXXXXX}

\acmJournal{JACM}
\acmVolume{37}
\acmNumber{4}
\acmArticle{111}
\acmMonth{8}

\begin{document}

\title{SURE: A Visualized Failure Indexing Approach using Program Memory Spectrum}

\author{Yi Song}
\authornote{Both authors contributed equally to this research.}
\email{yisong@whu.edu.cn}
\orcid{}
\author{Xihao Zhang}
\authornotemark[1]
\email{zhangxihao@whu.edu.cn}
\affiliation{%
  \institution{School of Computer Science, Wuhan University}
  \streetaddress{No. 299 Bayi Road, Wuchang District}
  \city{Wuhan}
  \state{Hubei}
  \country{China}
  \postcode{430072}
}

\author{Xiaoyuan Xie}
\authornote{Corresponding author.}
\affiliation{%
  \institution{School of Computer Science, Wuhan University}
  \streetaddress{No. 299 Bayi Road, Wuchang District}
  \city{Wuhan}
  \state{Hubei}
  \country{China}}
\email{xxie@whu.edu.cn}

\author{Songqiang Chen}
\affiliation{%
  \institution{The Hong Kong University of Science and Technology}
  \streetaddress{Clear Water Bay}
  \city{Hong Kong}
  \country{China}}
\email{i9s.chen@connect.ust.hk}

\author{Quanming Liu}
\affiliation{%
 \institution{School of Computer Science, Wuhan University}
 \streetaddress{No. 299 Bayi Road, Wuchang District}
 \city{Wuhan}
 \state{Hubei}
 \country{China}}
\email{liuquanming@whu.edu.cn}

\author{Ruizhi Gao}
\affiliation{%
  \institution{Sonos Inc.}
  \streetaddress{2 Ave de Lafayette}
  \city{Boston}
  \state{MA}
  \country{USA}}
\email{youtianzui.nju@gmail.com}

\renewcommand{\shortauthors}{Song, et al.}

\begin{abstract}
	Failure indexing is a longstanding crux in software testing and debugging, the goal of which is to automatically divide failures (e.g., failed test cases) into distinct groups according to the \syrevise{culprit} root causes, as such multiple faults residing in a faulty program can be isolated and thus be handled independently and simultaneously. The community of failure indexing has long been plagued by two challenges: 1) The effectiveness of division is still far from promising. Specifically, existing failure indexing techniques only employ a limited source of software run-time data, for example, code coverage, to be failure proximity and further divide them, which typically delivers unsatisfactory results. 2) The outcome can be hardly \syrevise{comprehensible}. Specifically, a developer who receives the division result is just aware of how all failures are divided, without knowing why they \syrevise{should be} divided the way they are. This leads to difficulties for developers to be convinced by the division result, \syrevise{which in turn affects the adoption of the results}. To tackle these two problems, in this paper, we propose SURE, a vi\textbf{SU}alized failu\textbf{R}e ind\textbf{E}xing approach using the program memory spectrum. We first collect the run-time memory information (i.e., variables’ names and values, as well as the depth of the stack frame) at several preset breakpoints during the execution of a failed test case, and transform the gathered memory information into a human-friendly image (called program memory spectrum, PMS). Then, any pair of PMS images that serve as proxies for two failures is fed to a trained Siamese convolutional neural network, to predict the likelihood of them being triggered by the same fault. Last, a clustering algorithm is adopted to divide all failures based on the mentioned likelihood. In the experiments, we use 30\% of the simulated faults to train the neural network, and use 70\% of the simulated faults as well as real-world faults to test. Results demonstrate the effectiveness of SURE: It achieves 101.20\% and 41.38\% improvements in faults number estimation, as well as 105.20\% and 35.53\% improvements in clustering, compared with the state-of-the-art technique in this field, in simulated and real-world environments, respectively. Moreover, we carry out a human study to quantitatively evaluate the \syrevise{comprehensibility} of PMS, revealing that this novel type of representation can help developers better \syrevise{comprehend} failure indexing results.
\end{abstract}






\maketitle

\section{Introduction}
\label{sect:introduction}

Due to the increments in the scale and complexity of modern software systems, faulty programs typically contain more than one fault~\cite{digiuseppe2015fault, gao2018research, jones2002visualization, wang2008software}. The co-existence of multiple faults has been shown to be harmful for fault localization tasks, that is, the existence of two or more faults will cause the localization of one or more of these faults to become harder~\cite{digiuseppe2015fault, callaghan2023improving}. A mainstream way to tackle the challenge of multi-fault localization is \textbf{\emph{parallel debugging}}\footnote{In parallel debugging, people generally consider faults that do not interfere with one another.}, in which a multi-fault localization task is decomposed into several sub-single-fault localization tasks by dividing\footnote{In the current field of parallel debugging, \emph{clustering} is typically utilized for such \emph{division}. Thus we use these two terms interchangeably in this paper.} all failures\footnote{In dynamic testing, \emph{failures} are typically \emph{failed\ test\ cases}. In this paper, we use these two terms interchangeably.} into distinct groups according to the \syrevise{culprit} faults~\cite{gao2019mseer,digiuseppe2012concept,golagha2019failure,pei2021dynamic,steimann2012improving}. Such a division process is generally referred to as \textbf{\emph{failure indexing}}, the goal of which is to isolate multiple faults into independent environments through the reorganization of failed test cases. Failure indexing has been proven to be diversely useful. For example, it can effectively alleviate the negative impact of multi-fault co-existence on fault localization by isolating faults. And for another example, the derived sub-tasks can be handled by different developers simultaneously, thus improving the efficiency of debugging~\cite{zakari2019community}.

For its promise, there are numerous works focusing on the topic of failure indexing~\cite{gao2019mseer,huang2013empirical,song2022comprehensive}. Despite the fact that failure indexing has been integrated into many approaches or tools, it is plagued by two longstanding challenges:

\begin{itemize}
	
	\item \textbf{The effectiveness of failure indexing is far from promising.} As mentioned above, the mission of failure indexing is to isolate multiple faults into separate environments by dividing failed test cases. It is obvious that an effective failure indexing process should satisfy two points: 1) The number of the derived fault-focused groups should be the same as the number of faults (i.e., \emph{correct faults number estimation}), and 2) Failures in the same group should be triggered by the same fault, and conversely, failures in different groups should have different root causes (i.e., \emph{promising clustering}). Many pioneering studies have pointed out that these two goals are difficult to achieve, because it is not easy to design an effective \textbf{\emph{failure proximity}} \syrevise{(comprising the fingerprinting function and the distance metric, we will introduce them in detail in Section~\ref{sect:background})}, which is the core of failure indexing~\cite{liu2008systematic}. The most prevalent and advanced failure \syrevise{proximities} to date \syrevise{are} the code coverage (CC)-based one and the statistical debugging (SD)-based one~\cite{huang2013empirical, liu2006failure}. However, they both rely on coverage information, \syrevise{which can be easily trapped and thus delivering unsatisfactory results in practice} (we will further describe this point in Section~\ref{subsect:fingerprintingfunction} and Section~\ref{sect:motivatingexample}).
	
	\iffalse A failure proximity comprises a fingerprinting function that extracts the signatures of failures, and the tailored distance metric that calculates the distance between failures based on the extracted signatures~\cite{}. \fi
	
\item \textbf{The outcome of failure indexing is hard to comprehend.} Given a collection of failed test cases, a failure indexing process will index them to their root causes and thus produce several fault-focused groups. However, developers who receive the result are only aware of how failures are divided, without knowing why they should be divided in the current manner. It is recognized that software engineering is a human-centric discipline~\cite{sommerville2011software}, the results provided by automated debugging techniques still require being \syrevise{comprehended} by the person who performs the following task~\cite{xie2016revisit,callaghan2023improving}. \syrevise{In fact, nearly thirty years ago, the comprehensibility of software debugging tasks has been considered by researchers. Specifically, to make developers better comprehend fault localization results, Agrawal et al. and Jones et al. used color to visually reveal the linkage between program statements and failed/passed runs. These works have been demonstrated to be effective in helping a user localize faults~\cite{agrawal1995fault, jones2002visualization}. Parnin and Orso also realized this problem and carried out a study, explicitly indicating that human factors played a critical role in software debugging tasks~\cite{parnin2011automated}. Moreover, Xie et al. suggested integrating comprehension assistance into automated debugging, because the results delivered by automated debugging techniques still require being comprehended by human developers who perform the following jobs~\cite{xie2016revisit}. In 2023, Souza et al. continue highlighting the importance of comprehension in software debugging tasks. They employed Jaguar~\cite{ribeiro2018jaguar}, an SBFL tool that visualizes fault localization results by coloring the suspicious elements from red (high) to green (low), and investigated to which extent such visualization provided hints for human developers to comprehend faulty behaviors~\cite{souza2023understanding}. The aforementioned representative works have garnered more than 2,700 citations. That is to say, comprehensibility has been recognized as an important topic in the field of software debugging. Seeing that failure indexing is an important class of tasks in software debugging, we can claim that \textbf{comprehension matters in failure indexing}}: only when human developers \syrevise{comprehend} the machine-produced failure indexing result, they will be convinced by the result and would like to further apply this result. Therefore, providing the failure indexing result and providing why the result is obtained are both important. However, to the best of our knowledge, existing studies only consider the former without focusing on the \syrevise{comprehensibility of failure indexing results}.

\end{itemize}

These two challenges are worth addressing because they are in line with two problems in practice, respectively: First, the effectiveness of failure indexing determines \emph{how good failure indexing will be}, and second, the \syrevise{comprehensibility} of the result determines \emph{whether the failure indexing result will be adopted by human developers.} In this paper, we propose SURE, a vi\textbf{SU}alized failu\textbf{R}e ind\textbf{E}xing approach based on the program memory spectrum for tackling the aforementioned challenges:

\begin{itemize}
	
	\item \textbf{To further improve the effectiveness of failure indexing,} we propose a novel type of failure proximity, namely, the Program Memory (PM)-based failure proximity. \syrevise{The shortcoming of using coverage to represent failures can be partly described by the PIE (Propagation, Infection, Execution) model~\cite{voas1992pie}, a classical theory in software testing and debugging. Specifically, the PIE model has demonstrated that a failure can be detected only if the fault\footnote{In this paper, we follow the practice of general fault localization to consider the non-omission fault.} infects \syrevise{the program's internal state}. However, with the coverage information only, it is very hard to explore the faulty internal state in depth during the program execution, because \textbf{being covered is a necessary but not sufficient condition for triggering a failure}. Thus, coverage cannot extract the signature of failures in deep insight. Based on this intuition, we conjecture that program internal dataflows (program memory in our context) can be a finer-grained failure representation when using coverage gives rise to unsatisfactory effectiveness, because program internal dataflows can effectively embody program internal states\footnote{This conjecture has been verified by the experiments in Section~\ref{sect:result}.}.} For the fingerprinting function, the PM-based failure proximity mines and utilizes the run-time memory information \syrevise{(i.e., variables' names, variables' values, and the depth of stack frame, \textbf{deeper insight to explore programs' internal state than coverage})} collected during the execution of a failed test case to represent it. Specifically, we first employ Spectrum-Based Fault Localization (SBFL) techniques to determine several highly-suspicious program statements, and set these statements as breakpoints. And then, we collect the run-time memory information (i.e., variables' names and values, as well as the depth of the stack frame) at the preset breakpoints during executing a failed test case. We further reorganize the collected run-time memory information into Program Memory Spectrum (PMS),	which is in the form of an image, to serve as the proxy for a failed test case. And for the distance metric, we train a Siamese convolutional neural network to predict the distance between a pair of PMS images (i.e., a pair of failed test cases), which reflects the likelihood that they are triggered by the same fault.
	
	\item \textbf{To make failure indexing results \syrevise{comprehensible} to human developers,} we borrow the idea of \syrevise{Agrawal et al.'s work and} Jones et al.'s work to make the failure indexing result visualized. In their work, they used color to visually map program statements in test suite executions, and help developers pinpoint faulty statements intuitively~\cite{agrawal1995fault, jones2002visualization}. In this paper, we design a novel algorithm to reorganize the collected run-time memory information (\syrevise{which is not easy for human developers to recognize}) into PMS images (which are in a human-friendly form). Specifically, for a set of run-time memory information collected during running a failed test case, we first convert variables' names and values into the numeric form according to the Ascii code of characters, and then regard variables' names, variables' values, and the depth of the stack frame as the three channels of RGB \syrevise{(we will further introduce RGB in Section~\ref{subsect:generatePMS})}, to represent this failure in the form of PMS. As such, each failed test case is represented as a visualized image, so human developers can be easily aware of why all failures are divided \syrevise{the way they are} when they receive the failure indexing result.
\end{itemize}

In the experiments, we obtain four C projects, Flex, Grep, Gzip, and Sed from the Software Infrastructure Repository (SIR)~\cite{do2005supporting}, and employ a mutation tool to inject simulated faults into the projects to generate faulty versions that contain one, two, three, four or five bugs (also referred to as 1-bug, 2-bug, 3-bug, 4-bug, and 5-bug faulty versions\syrevise{, respectively}). Besides, we also use five Java projects, Chart, Closure, Lang, Math, and Time from Defects4J~\cite{just2014defects4j}, as our benchmarks, and gather 1 $\sim$ 5-bug real-world faulty versions according to the test case transplantation strategy proposed by An et al.~\cite{an2021searching}. In total, we generate 1,000 C simulated faulty versions and gather 100 Java real-world faulty versions. We use only 30\% of the simulated faulty versions to train, and use 70\% of the simulated faulty versions as well as real-world faulty versions to test. Experimental results significantly demonstrate the promise of the proposed approach: SURE can achieve 101.20\% and 41.38\% improvements in faults number estimation, as well as 105.20\% and 35.53\% improvements in clustering, compared with the state-of-the-art technique in the field, in simulated and real-world environments, respectively.

Moreover, to quantitatively measure to which extent the proposed visualized representation of failed test cases (i.e., PMS images) can be \syrevise{comprehensible} to human developers, we conduct a human study involving 15 graduate students in Computer Science. The results show that compared with the most advanced and prevalent failure representation to date (i.e., the ranking-based and coverage-based ones), the representation of PMS images is more \syrevise{comprehensible} and cost-effective: The \syrevise{comprehensibility} of PMS images is 105.26\% higher than the two baselines, while the time cost is only \syrevise{15.35\%} and \syrevise{14.87\%} of the two baselines, respectively.

The main contributions of this paper are as follows:

\begin{enumerate}
	
	\item \textbf{A novel failure indexing approach.} We propose a visualized failure indexing approach, SURE, which utilizes the program memory spectrum to represent failures and uses Siamese convolutional neural networks to measure the distance between failures.
	
	\vspace{3pt}
	
	\item \textbf{A visualized representation of failed test cases.} The program memory spectrum is a visualized representation of failed test cases, which is in a human-friendly form (i.e., images) and thus can make failure indexing outcomes more \syrevise{comprehensible} for human developers.
	
	\vspace{3pt}
	
	\item \textbf{A comprehensive evaluation.} We use both simulated and real-world faults in the experiments for more robust evaluation. It is worth mentioning that the training phase is performed on a small set of simulated faults while the test is conducted on a larger set of simulated faults and real-world faults. We also carry out a human study to quantitatively demonstrate the \syrevise{comprehensibility} of the proposed PMS.

\end{enumerate}

The remainder of this paper is organized as follows: Section~\ref{sect:background} introduces the background knowledge and emphasizes the motivation of this paper. Section~\ref{sect:motivatingexample} provides a motivating example. Section~\ref{sect:approach} describes the technique details of SURE, and gives a running example to facilitate comprehension. Section~\ref{sect:setup} lists the research questions, datasets, evaluation metrics, etc. Section~\ref{sect:result} analyzes the experimental results. Section~\ref{sect:discussion} discusses some interesting topics. Section~\ref{sect:threats} is about the threats to validity. Section~\ref{sect:relatedwork} reports related works. Conclusions and directions for future work are proposed in Section~\ref{sect:conclusion}.

\section{Background}
\label{sect:background}

\subsection{Failure Indexing}
\label{subsect:failureindexing}

Parallel debugging is a well-recognized strategy for multi-fault localization, one of the most tedious and time-consuming problems in software testing and debugging~\cite{zakari2020multiple,wong2016survey,digiuseppe2011influence,keller2017critical,wang2019fuzzy}. The core of parallel debugging is to correctly divide all failures into distinct groups according to their root causes, i.e., failure indexing, thus multiple faults can be localized in isolated environments. It is obvious that failure indexing will directly determine the effectiveness of parallel debugging: the more promising failure indexing is, the better parallel debugging~\cite{Li2022review,wu2020fatoc,jones2007debugging}.

A collection of research has pointed out that failure indexing is indeed a difficult task. Failure indexing typically involves three components: A fingerprinting function that represents failures in a mathematical and structured form, a distance metric that calculates the distance between the proxies for failures, and a clustering algorithm that divides all failures into several groups based on the calculated distance information. Previous studies have made it clear that there is no clustering technique that is universally applicable in uncovering the variety of structures present in multidimensional data sets~\cite{jain1999data}. Thus, the most essential factors of failure indexing are the fingerprinting function and the distance metric \syrevise{(these to components are called \emph{\textbf{failure proximity}}).}

\subsection{Fingerprinting Function}
\label{subsect:fingerprintingfunction}

The directly available information regarding \syrevise{failed test cases to be indexed} is twofold: the input data and the testing result (i.e., \emph{failed}), which is too limited to effectively differentiate them. To tackle this problem, researchers typically \syrevise{design fingerprinting functions to mine dynamic information at run-time, so as to} further extract the signature of failed test cases and use such data to represent failed test cases~\cite{dickinson2001finding,podgurski2003automated,liu2006failure}.

Among off-the-shelf fingerprinting functions, 
\iffalse
off-the-shelf fingerprinting functions main include the failure point-based~\cite{van2018semantic}, the stack trace-based~\cite{gu2019does}, the predicate evaluation-based~\cite{jones2007debugging}, the dynamic slicing-based~\cite{liu2008systematic}, the code coverage-based~\cite{huang2013empirical,hogerle2014more,digiuseppe2012software}, and the statistical debugging-based ones~\cite{gao2019mseer, jones2007debugging,li2022preliminary,song2022comprehensive}.
\fi
the code coverage-based strategy is the most prevalent currently~\cite{huang2013empirical,hogerle2014more,digiuseppe2012software}, while the statistical debugging-based strategy is the most sophisticated and advanced~\cite{gao2019mseer, jones2007debugging,li2022preliminary,song2022comprehensive}. Here we use a faulty program containing $l$ executable statements as an example. The code coverage (CC)-based fingerprinting function extracts the signature of failures from the execution path of failed test cases. It will represent failed test case $f$ as a numerical vector of length $l$. There are two variants of the CC-based fingerprinting function: $Cov_{hit}$ and $Cov_{count}$. As for the former, the $i^{\rm{th}}$ element of the numerical vector that represents $f$ will be set to 1 if $f$ covers the $i^{\rm{th}}$ statement $s_i$, and 0 otherwise. And as for the latter, the $i^{\rm{th}}$ element of the numerical vector that represents $f$ will be set to the execution frequency of $s_i$ if $f$ covers $s_i$, and 0 otherwise. The statistical debugging (SD)-based fingerprinting function extracts the signature of failures from the suspiciousness ranking list of program statements. Specifically, a failed test case $f$ is first combined with passed test cases $T_S$, then the test suite $f \cup T_S$ is executed on the faulty program and the coverage is collected simultaneously. Later, a spectrum-based fault localization (SBFL) technique is employed to calculate the suspiciousness value of each program statement based on coverage information~\cite{xie2013theoretical,yoo2017human,naish2011model}, and produce a ranking list in which $l$ program statements are descendingly ordered by their suspiciousness. As such, $f$ can be represented as this ranking list of length $l$.

Despite the promise of the CC and the SD-based strategies, they share a common bottleneck: only relying on program coverage. Specifically, the intuition of the CC-based strategy is that failures triggered by the same fault should also have the same code coverage, and vice versa~\cite{yoon2021enhancing}. Thus, it creates numeric vectors that reflect the coverage information during the execution to represent failed test cases. The intuition of the SD-based strategy is that failures triggered by the same fault should target the same fault location, and vice versa~\cite{liu2008systematic, liu2006failure}. Thus, it incorporates SBFL techniques to further analyze raw coverage information, to convert it to a suspiciousness ranking list that suggests the fault location and thus can represent failed test cases. Obviously, both CC and SD purely rely on coverage information, which cannot work well if the coverage of the failures triggered by different faults is the same \syrevise{(we will exemplify such a situation in Section~\ref{sect:motivatingexample})}. That is to say, \textbf{even the most prevalent and advanced fingerprinting functions to date are still not able to deliver satisfactory results.}

\subsection{Distance Metric}
\label{subsect:distancemetric}

In failure proximity, the design of the distance metric is not an independent task, because it is tightly related to the form of the fingerprinting function. For example, the CC-based strategy represents a failed test case as a numerical vector and thus typically uses the Euclidean distance metric~\cite{vijaymeena2016survey}, because the Euclidean distance metric is a suitable and cheap way to handle the similarity measurement between such numerical vectors~\cite{huang2013empirical}. For another example, the SD-based strategy represents a failed test case as a ranking list and thus typically uses the Kendall tau distance metric~\cite{kendall1948rank}, because the Kendall tau distance metric can properly measure how similar two ranking lists are by counting the number of pairwise disagreements between them~\cite{gao2019mseer}. From these, we can see that the distance metric is inseparable from the fingerprinting function: \textbf{only by designing a well-tailored distance metric can the signature of failures extracted by the fingerprinting function be fully exploited.}

\section{Motivating Example}
\label{sect:motivatingexample}

We use a motivating example in Table~\ref{tab:motivatingexample} to reveal the shortcoming of the CC-based and SD-based failure proximities mentioned above. This toy program containing 17 statements \syrevise{($s_1$, $s_2$, ..., $s_{17}$)} is used to identify and replace certain words in the input string, and then output the modified string and the log message. Specifically, if an input string contains ``\emph{wordNone}'' or ``\emph{wordNtwo}'', these two words will be replaced with ``\emph{*1*}'' and ``\emph{*2*}'', respectively. The log message records the operation of the program, for example, ``\emph{wordNone recognized}'', ``\emph{wordNtwo recognized}'', ``\emph{both pattern recognized}'', and ``\emph{pass}''. In this program, statements $s_8$ and $s_{16}$ each contain a fault.

Given a test suite containing 12 test cases: $t_1$ = ``\emph{speak\ wordNone}'', $t_2$ = ``\emph{wordNone}'', $t_3$ = ``\emph{wordNonecontained}'', $t_4$ = ``\emph{wwwwordNoneeee}'', $t_5$ = ``\emph{has\ wordNtwo}'', $t_6$ = ``\emph{wordNtwo}'', $t_7$ = ``'', $t_8$ = ``\emph{midd*1*le}'', $t_9$ = ``\emph{*1*2*}'', $t_{10}$ = ``\emph{a normal sentence}'', $t_{11}$ = ``\emph{wordnonewordNtw}'', and $t_{12}$ = ``\emph{wordNone and wordNtwo}''. The coverage information of them is given in Table~\ref{tab:motivatingexample}, where a dot in the cell [$t_i$, $s_j$] indicates that $t_i$ covers $s_j$ ($i$ = 1, 2, ..., 12, and $j$ = 1, 2, ..., 17). Among these test cases, $t_1$ $\sim$ $t_6$ are failed test cases due to the inconsistency between the actual output and the expected output ($t_1$ $\sim$ $t_6$ can be referred to as $f_1$ $\sim$ $f_6$, respectively). More concretely, $f_1$ $\sim$ $f_4$ are triggered by $Fault_1$, while $f_5$ and $f_6$ are triggered by $Fault_2$. The remaining six test cases, i.e., $t_7$ $\sim$ $t_{12}$, pass the test (they can be referred to as $T_S$).

\begin{table}\footnotesize
	\caption{A motivating example}
	\label{tab:motivatingexample}
	\begin{tabular}{cp{150pt}cccccccccccc}
		\toprule
		\textbf{S} & \textbf{Program} & \bm{$t_1$} & \bm{$t_2$} & \bm{$t_3$} & \bm{$t_4$} & \bm{$t_5$} & \bm{$t_6$}& \bm{$t_7$ }& \bm{$t_8$} & \bm{$t_9$} & \bm{$ t_{10} $}& \bm{$t_{11} $} & \bm{$t_{12} $}	\\
		\midrule
		
		$s_1$ & public static String process(String s)\{ & $\bigcdot$ & $\bigcdot$ & $\bigcdot$ & $\bigcdot$ & $\bigcdot$ & $\bigcdot$ & $\bigcdot$ & $\bigcdot$ & $\bigcdot$ & $\bigcdot$ & $\bigcdot$ & $\bigcdot$ \\
		$s_2$ & \qquad if(s.contains(``*1*'') \textbar{}\textbar{} s.contains(``*2*''))\{ & $\bigcdot$ & $\bigcdot$ & $\bigcdot$ & $\bigcdot$ & $\bigcdot$ & $\bigcdot$ & $\bigcdot$ & $\bigcdot$ & $\bigcdot$ & $\bigcdot$ & $\bigcdot$ & $\bigcdot$ \\
		$s_3$ & \qquad\qquad return "";\}  & & & & & & &  & $\bigcdot$ & $\bigcdot$ & &  &  \\
		$s_4$ & \qquad int sign = 0; & $\bigcdot$ & $\bigcdot$ & $\bigcdot$ & $\bigcdot$ & $\bigcdot$ & $\bigcdot$ & $\bigcdot$ & &  & $\bigcdot$ & $\bigcdot$ & $\bigcdot$\\
		$s_5$ & \qquad int sum\_1 = 0;  & $\bigcdot$ & $\bigcdot$ & $\bigcdot$ & $\bigcdot$ & $\bigcdot$ & $\bigcdot$ & $\bigcdot$ & &  & $\bigcdot$ & $\bigcdot$ & $\bigcdot$ \\
		$s_6$ & \qquad sum\_1 = s.contains(``wordNone'') ? 1 : 0; & $\bigcdot$ & $\bigcdot$ & $\bigcdot$ & $\bigcdot$ & $\bigcdot$ & $\bigcdot$ & $\bigcdot$ & &  & $\bigcdot$ & $\bigcdot$ & $\bigcdot$ \\
		$s_7$ &  \qquad sign += sum\_1; & $\bigcdot$ & $\bigcdot$ & $\bigcdot$ & $\bigcdot$ & $\bigcdot$ & $\bigcdot$ & $\bigcdot$ & &  & $\bigcdot$ & $\bigcdot$ & $\bigcdot$ \\
		$s_8$ & \begin{tabular}[c]{@{}l@{}}\qquad s = s.replaceAll(``wordNone'', ``?1?''); \\\qquad \comm{\qquad //Fault1: ``?1?'' should be ``*1*''}\end{tabular} & $\bigcdot$ & $\bigcdot$ & $\bigcdot$ & $\bigcdot$ & $\bigcdot$ & $\bigcdot$ & $\bigcdot$ & &  & $\bigcdot$ & $\bigcdot$ & $\bigcdot$  \\
		$s_9$ & \qquad int sum\_2 = 0; & $\bigcdot$ & $\bigcdot$ & $\bigcdot$ & $\bigcdot$ & $\bigcdot$ & $\bigcdot$ & $\bigcdot$ & &  & $\bigcdot$ & $\bigcdot$ & $\bigcdot$ \\
		$s_{10}$ & \qquad sum\_2 = s.contains(``wordNtwo'') ? 2 : 0; & $\bigcdot$ & $\bigcdot$ & $\bigcdot$ & $\bigcdot$ & $\bigcdot$ & $\bigcdot$ & $\bigcdot$ & &  & $\bigcdot$ & $\bigcdot$ & $\bigcdot$  \\
		$s_{11}$ & \qquad sign += sum\_2; & $\bigcdot$ & $\bigcdot$ & $\bigcdot$ & $\bigcdot$ & $\bigcdot$ & $\bigcdot$ & $\bigcdot$ & &  & $\bigcdot$ & $\bigcdot$ & $\bigcdot$ \\
		$s_{12}$ & \qquad s = s.replaceAll(``wordNtwo'', ``*2*''); & $\bigcdot$ & $\bigcdot$ & $\bigcdot$ & $\bigcdot$ & $\bigcdot$ & $\bigcdot$ & $\bigcdot$ & &  & $\bigcdot$ & $\bigcdot$ & $\bigcdot$ \\
		$s_{13}$ & \qquad if(sign == 3)\{ & $\bigcdot$ & $\bigcdot$ & $\bigcdot$ & $\bigcdot$ & $\bigcdot$ & $\bigcdot$ & $\bigcdot$ & &  & $\bigcdot$ & $\bigcdot$ & $\bigcdot$ \\
		$s_{14}$ & \qquad\qquad return ``both pattern recognized'';\}  & & & & & & & & & & & & $\bigcdot$  \\
		$s_{15}$ & \begin{tabular}[c]{@{}l@{}}\qquad String msg = sign == 1 ? \\ \qquad ``wordNone recognized'' : ``pass'';\end{tabular} & $\bigcdot$ & $\bigcdot$ & $\bigcdot$ & $\bigcdot$ & $\bigcdot$ & $\bigcdot$ & $\bigcdot$ &  & & $\bigcdot$ & $\bigcdot$ &  \\
		$s_{16}$ & \begin{tabular}[c]{@{}l@{}}\qquad msg = sign \textgreater{} 2 ? \\ \qquad``wordNtwo recognized'' : msg; \\ \qquad \comm{\qquad//Fault2: ``>2'' should be ``==2''} \end{tabular} & $\bigcdot$ & $\bigcdot$ & $\bigcdot$ & $\bigcdot$ & $\bigcdot$ & $\bigcdot$ & $\bigcdot$ &  & & $\bigcdot$ & $\bigcdot$ &  \\
		
		\vspace{3pt}
		
		$s_{17}$ & \qquad return s + ``//'' + msg;\} & $\bigcdot$ & $\bigcdot$ & $\bigcdot$ & $\bigcdot$ & $\bigcdot$ & $\bigcdot$ & $\bigcdot$ &  & & $\bigcdot$ & $\bigcdot$ & \\
		& \qquad\qquad\qquad \textbf{Result} & \textbf{F} & \textbf{F}  & \textbf{F}  & \textbf{F}  & \textbf{F}  & \textbf{F}  & \textbf{S}  & \textbf{S}  & \textbf{S}  & \textbf{S}  & \textbf{S}  & \textbf{S} 
		\\
		
		\bottomrule
	\end{tabular}
\end{table}

\begin{table}[]
	\centering
	\caption{\label{tab:motivatingexampleproxy} \syrevise{The representation for $f_2 \sim f_6$}}
	\begin{tabular}{p{3.7cm}<{\centering}p{6cm}<{\centering}}
		\hline
		 \textbf{Failure Proximity} &  \textbf{Representation}   \\ \hline
		 The CC-based &  [1, 1, 0, 1, 1, 1, 1, 1, 1, 1, 1, 1, 1, 0, 1, 1, 1]   \\ 
		 The SD-based &  [1, 1, 3, 4, 4, 4, 4, 4, 4, 4, 4, 4, 4, 4, 15, 15, 15]  \\

		\hline	
	\end{tabular}
\end{table}

As mentioned in Section~\ref{subsect:fingerprintingfunction}, the CC-based failure proximity represents a failure as a code coverage vector, i.e.,  the corresponding column in Table~\ref{tab:motivatingexample}. For example, $f_1$ covers all program statements except $s_3$ and $s_{14}$, thus, as for $Cov_{hit}$, $f_1$ can be represented as a binary vector of length 17: [1, 1, 0, 1, 1, 1, 1, 1, 1, 1, 1, 1, 1, 0, 1, 1, 1]. \syrevise{By observing Table~\ref{tab:motivatingexample}, we can find that the CC-based representation for the other five failed test cases ($f_2 \sim f_6$) is the same as that for $f_1$, as shown in the second row of Table~\ref{tab:motivatingexampleproxy}. And as for $Cov_{count}$, the number ``1'' in the binary vector will be replaced with the execution frequency, which does not help to differentiate the six failures either. Thus, neither $Cov_{hit}$ nor $Cov_{count}$ can effectively perform failure indexing in this example, i.e., the scenario where failures caused by different faults have the same execution coverage.}

Also recall the description in Section~\ref{subsect:fingerprintingfunction}, the SD-based failure proximity represents a failed test case as a suspiciousness ranking list of program statements based on this failed test case and passed test cases $T_S$. For example, \syrevise{to represent $f_1$, }we can combine $f_1$ with $T_S$ to form a new test suite, $f_1$ $\cup$ $T_S$, run this test suite on the program, and collect the coverage information. The coverage will be delivered to an SBFL technique to \syrevise{calculate risk values of being faulty for each program statement, and then }produce a suspiciousness ranking list \syrevise{by descendingly ordering statements by their risk values}. Here we employ a recognized SBFL technique, GP03~\cite{yoo2012evolving}, whose expression is given in Formula~\ref{equ:gp03}, to complete this process\footnote{\syrevise{SBFL techniques typically involve four spectrum notations: $N_{CF}$ and $N_{CS}$ represent the number of test cases that execute the statement and return the testing result of failed or passed, respectively, $N_{UF}$ and $N_{US}$ represent the number of test cases that do not execute it and return the testing result of failed or passed, respectively.}}.

\vspace{5pt}
\begin{equation}
	\label{equ:gp03}
	Suspiciousness_{GP03} = \sqrt{| N_{C F} ^ 2 - \sqrt{N_{C S}} |}.
\end{equation}
\vspace{5pt}

The generated suspiciousness ranking list is [1, 1, 3, 4, 4, 4, 4, 4, 4, 4, 4, 4, 4, 4, 15, 15, 15]\footnote{\syrevise{In light of the experience of previous works~\cite{huang2013empirical, gao2019mseer, xu2011ties}, if several statements with the same suspiciousness form a tie, the rankings of all statements in the tie will be set to the beginning position of this tie.}}, which will serve as the proxy for $f_1$. However, the SD-based representation for the other five failed test cases ($f_2 \sim f_6$) is the same as that for $f_1$, as shown in the third row of Table~\ref{tab:motivatingexampleproxy}. That is, despite the incorporation of SBFL techniques, the SD-based failure proximity can still not differentiate the six failed test cases, because the SD-based strategy essentially relies on program coverage as well, while the six failed test cases triggered by different faults here exhibit the same execution coverage.

\syrevise{That is to say, when failures caused by different faults have identical coverage, even the most widespread and state-of-the-art failure proximities are still not enough to deliver promising outcomes.} This situation, is, unfortunately, very common in practice, as a prestigious work in the field of failure indexing pointed out previously:
\vspace{6pt}
\begin{center}
	\begin{tcolorbox}[colback=gray!15,%gray background
		colframe=black,% black frame colour
		width=10cm,% Use 8cm total width,
		height = 1cm,
		arc=2mm, auto outer arc,
		boxrule=0.7pt,
		]
		{\vspace{-0.18cm}\emph{\syrevise{``A significant portion of execution profiles would be the same even if these failures are due to different faults''}}~\cite{liu2008systematic}.}
	\end{tcolorbox}
\end{center}
\vspace{6pt}
\syrevise{Thus, developing a better failure proximity is of great significance.}

Moreover, neither the CC-based nor the SD-based failure proximity represents failures in a human-friendly form, it is hard for human developers to \syrevise{comprehend} the failure indexing result based on such forms of failure representation. Specifically, whether a numerical vector in the CC-based strategy or a ranking list in the SD-based strategy, its length is equal to the number of executable statements of the faulty program. Unlike our motivating example, in real-world debugging, a faulty program can easily have tens of thousands of statements. As a consequence of which, the proxy for a failure will be a numerical vector or a ranking list of great length. It is obvious that in such a scenario, given failure indexing results, human developers are hard to \syrevise{comprehend} why these numerical vectors or ranking lists should be clustered into the same (or different) group(s).

This motivating example confirms the challenges we point out in Section~\ref{sect:introduction}, namely, 1) The effectiveness of failure indexing is far from promising, and 2) The outcome of failure indexing is hard to \syrevise{comprehend for human developers}. In this paper, we propose SURE, a novel failure indexing approach based on the program memory spectrum, which is more effective than the state-of-the-art failure indexing technique to date. Program memory spectrum is in the form of visualized images, human developers can easily \syrevise{comprehend} the failure indexing results given such human-friendly failure representation.

\section{Approach}
\label{sect:approach}

\begin{figure}[t]
	\centering
	\includegraphics[width=\linewidth]{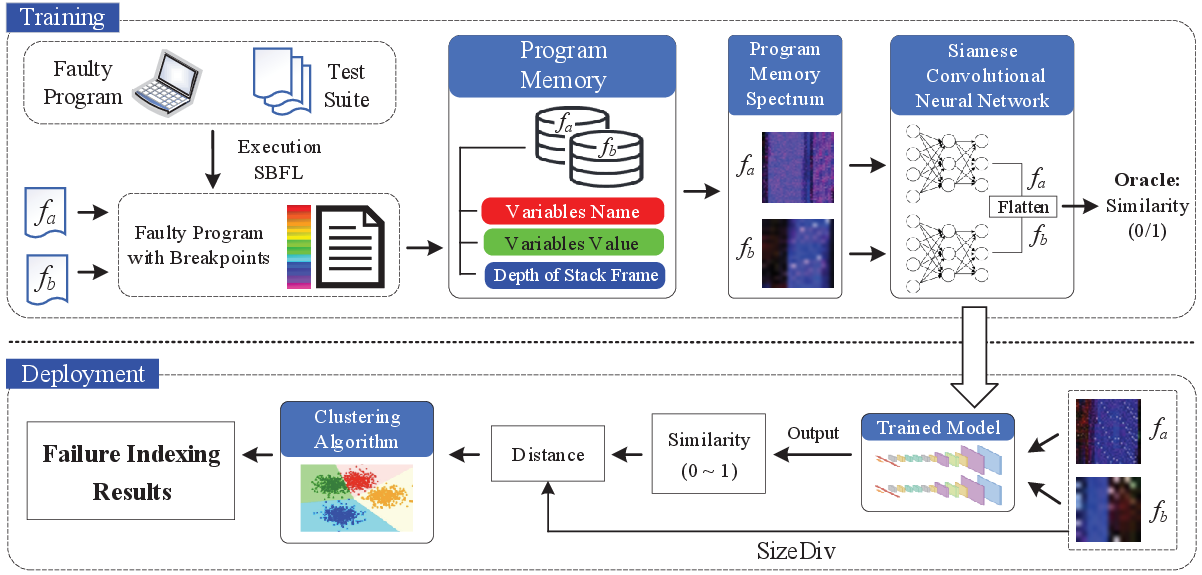}
	\caption{Overview of SURE}
	\label{fig:overview}
\end{figure}

\syrevise{The program memory-based failure proximity utilizes the run-time memory information to represent failures, and measures the distance between failures based on the characteristics of memory information. Based on this description,} we propose our approach, SURE. The overview of SURE is depicted in Figure~\ref{fig:overview}. For a faulty program under test and a given test suite containing multiple failed test cases, SURE performs the following steps to achieve failure indexing:

\begin{enumerate}
	\item \textbf{Breakpoint determination.} we first employ an SBFL technique to obtain the suspiciousness of all program statements, and then set the Top-$x$\% riskiest statements as the breakpoints (the determination of the value of $x$ will be investigated in the first research question of Section~\ref{subsect:rq}).
	
	\item \textbf{Collection of run-time program memory.} For a failed test case, we run it on the faulty program, and collect the run-time memory information at the preset breakpoints. Specifically, the run-time memory information we collect includes variables' names, variables' values, and the depth of the stack frame.
	
	\item \textbf{Generation of program memory spectrum.} The collected memory information will be converted to program memory spectrum (PMS), which is in the form of visualized images in a human-friendly way.
	
	\item \textbf{Model training.} For any pair of PMS images (i.e., any pair of failed test cases), we will assign a label to it: If these two failed test cases are triggered by the same fault, the label is 1, indicating that the similarity between them should be 1. Otherwise, if these two failed test cases have different root causes, the label is 0. Then, we feed such labeled pairs of PMS images to a Siamese convolutional neural network to train.
	
	\item \textbf{Prediction.} Once the model is trained, it can be used to handle the prediction of the similarity between two unseen PMS images.
	
	\item \textbf{Distance calculation.} For the value of the similarity between a pair of failed test cases, we first subtract it from 1, and then multiply this difference by a parameter, $SizeDiv$, as the final distance. $SizeDiv$ reflects the divergence between the sizes of two PMS images.
	
	\item \textbf{Clustering.} The clustering algorithm will receive the distances between any pair of failed test cases, and based on which produces several clusters of failed test cases.  
	
\end{enumerate}

We detailedly introduce the above seven steps of SURE in Section~\ref{subsect:breakpointdetermination} $\sim$ Section~\ref{subsect:clustering}, respectively. Moreover, we give a running example to facilitate readers' comprehension in Section~\ref{subsect:runningexample}.

\subsection{Breakpoint determination}
\label{subsect:breakpointdetermination}

Given a faulty program $P$ containing $l$ statements, a test suite $T$ comprising failed test cases $T_F$ and passed test cases $T_S$ ($T_F$ $\cup$ $T_S$ = $T$ and $T_F$ $\cap$ $T_S$ = $\varnothing$). We employ an SBFL technique to calculate the suspiciousness of $l$ program statements of $P$ based on $T$, and rank all statements in descending order of suspiciousness. Then, we determine the Top-$x$\% riskiest statements as the breakpoints (the value of $x$ will be investigated in the first research question of Section~\ref{subsect:rq}). As such, we can get $q$ breakpoints, $bp_1$, $bp_2$, ... $bp_j$, ..., $bp_q$, where the value of $q$ is $\lfloor$ $l$ $\times$ $x$\% $\rfloor$, and $bp_j$ is the breakpoint ranked $j^{\rm{th}}$ in suspiciousness.

The intuition behind this strategy is that the statements with higher risk values are more likely to be faulty, and run-time memory information gathered at these positions could have a  stronger capability to reveal faults, thus can contribute more to representing failures.

\syrevise{As a reminder, SBFL is not required to determine the exact position of faults in this process. We simply need to identify a collection of highly-risky program statements, which are sufficient to specify breakpoints that can monitor the execution of abnormal programs.}

\subsection{Collection of run-time program memory}
\label{subsect:collectionmemory}

For the $i^{\rm{th}}$ failed test case in $T_F$, $f_i$ $\in$ $T_F$, executing it on $P$ and gathering the run-time memory information at each breakpoint. Specifically, the collected memory information of $f_i$ is:
\begin{equation}
\label{equ:mii}
MI_i=\left\{b p_1^{\prime}: V_1^i, b p_2^{\prime}: V_2^i, ..., b p_j^{\prime}: V_j^i, ..., b p_q^{\prime}: V_q^i\right\},
\end{equation}
where $V^i_j$ is the memory information collected at $bp_j^{\prime}$ during executing $f_i$. Notice the difference between $bp_j^{\prime}$ here and $bp_j$ in Section~\ref{subsect:breakpointdetermination}: $bp_j^{\prime}$ is the $j^{\rm{th}}$ breakpoint to be executed during the actual running of $f_i$. \syrevise{This arrangement integrates control flows information into our method, because program control flows are also helpful for failure representation to an extent}. Supposing that there are $m_{ij}$ variables queried at $bp_j^{\prime}$ during the running of $f_i$, $V^i_j$ comprises three \syrevise{lists} with length $m_{ij}$:  ${V_{name}}^i_j$,  ${V_{value}}^i_j$, and ${V_{depth}}^i_j$:

${V_{name}}^i_j$ contains the names of the $m_{ij}$ variables queried at $bp_j^{\prime}$ during running $f_i$:
\begin{equation}
	\label{equ:vnameij}
    {V_{name}}^i_j = [name_1, name_2, ..., name_{m_{ij}}] ,
\end{equation}

${V_{value}}^i_j$ contains the values of these $m_{ij}$ variables\footnote{In our experiments, we find that regarding a variable's value as a string is beneficial for distance measurement. For some special types of variables, further action will be taken. For example, for a pointer, we will further index its value by address.}:
\begin{equation}
	\label{equ:vvalueij}
	{V_{value}}^i_j = [value_1, value_2, ..., value_{m_{ij}}] ,
\end{equation}

${V_{depth}}^i_j$ contains the depth of the stack frame of these $m_{ij}$ variables:
\begin{equation}
	\label{equ:vdepthij}
	{V_{depth}}^i_j = [depth_1, depth_2, ..., depth_{m_{ij}}].
\end{equation}

If we are from the perspective of a failed test case, that is, considering all breakpoints together, we can integrate the memory information collected at all breakpoints into a hunk:

\begin{equation}
	\label{equ:vnamei}
	V_{name}^i = {V_{name}}^i_1 \concat {V_{name}}^i_2 \concat ... \concat {V_{name}}^i_j \concat ... \concat {V_{name}}^i_q\ ,
\end{equation}

\begin{equation}
	\label{equ:vvaluei}
	V_{value}^i = {V_{value}}^i_1 \concat {V_{value}}^i_2 \concat ... \concat {V_{value}}^i_j \concat ... \concat {V_{value}}^i_q\ ,
\end{equation}

\begin{equation}
	\label{equ:vdepthi}
	V_{depth}^i = {V_{depth}}^i_1 \concat {V_{depth}}^i_2 \concat ... \concat {V_{depth}}^i_j \concat ... \concat {V_{depth}}^i_q\,
\end{equation}

\syrevise{where $\concat$ denotes the append operation between two lists.} We use $m_i$ to denote the length of $V_{name}^i$, $V_{value}^i$, and $V_{depth}^i $, which is the number of variables queried at all breakpoints during the running of $f_i$:

\begin{equation}
	\label{equ:mi}
	m_i = \left|  V_{name}^i  \right|= \left|  V_{value}^i  \right| = \left|  V_{depth}^i  \right| = m_{i1} + m_{i2} + ... + m_{ij} + ... + m_{iq}
\end{equation}

\syrevise{During the execution of a failed test case, when the program stops at a breakpoint, only the memory data at the position before the execution of that breakpoint can be collected. To ensure that memory information is gathered regarding the predetermined positions produced by SBFL, when faulty programs stop at each of the breakpoints, we continue executing a further step and then carry out the collection operation. Besides, if a program statement is executed for more than once, we collect variables' values when the execution is completed, since the latest value reflects the entire accumulation of the execution. As a reminder, for the sake of cost saving, if a program statement contains function calls, we concentrate on the original location of the preset breakpoint rather than iteratively navigating to the callee to query memory information (Experiments in Section~\ref{sect:result} demonstrate that this strategy is good enough for failure indexing).}

\subsection{Generation of program memory spectrum}
\label{subsect:generatePMS}

For each $f_i$ $\in$ $T_F$, we have collected the run-time memory information regarding it, i.e., $MI_i$. Now we introduce based on $MI_i$, how to generate the corresponding program memory spectrum $PMS_i$.

\begin{figure}[t]
	\centering
	\includegraphics[width=\linewidth]{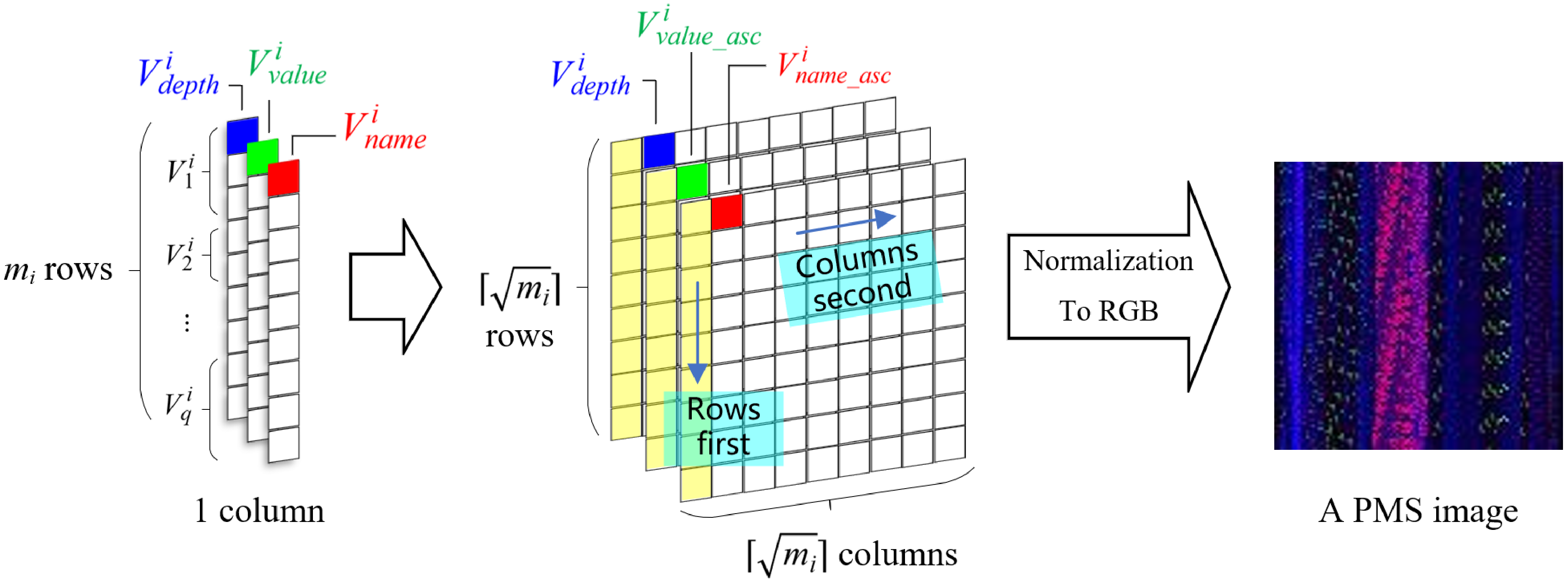}
	\caption{Workflow of PMS images generation}
	\label{fig:PMSGeneration}
\end{figure}

The workflow of the generation of PMS images is illustrated in Figure~\ref{fig:PMSGeneration}. Specifically, $MI_i$ is a matrix with the shape of (1 $\times$ $m_i$ $\times$ 3), as shown in \textbf{the leftmost sub-figure of Figure~\ref{fig:PMSGeneration}}, where $m_i$ is the number of variables queried at all breakpoints during the running of $f_i$, and ``3'' indicates the three dimensions: variables' names, variables' values, and the depth of the stack frame. The arrangement of this (1 $\times$ $m_i$ $\times$ 3) matrix is in the order of execution of breakpoints. Specifically, $V^i_1$, i.e., the run-time memory information collected at the first executed breakpoint during running $f_i$, is at the top. Similarly, $V^i_q$, i.e., the run-time memory information collected at the last executed breakpoint (there are $q$ breakpoints in total) during running $f_i$, is at the bottom.

In practical development, the value of $m_i$ is generally very large, thus the shape of this (1 $\times$ $m_i$ $\times$ 3) matrix is quite unbalanced. We reshape this matrix to a ($\lceil \sqrt{m_i} \rceil$ $\times$ $\lceil \sqrt{m_i} \rceil$ $\times$ 3) matrix. The principle of this reshaping process is ``row first, column second''. Specifically, we take the first $\lceil \sqrt{m_i} \rceil$ elements from top to bottom in the (1 $\times$ $m_i$ $\times$ 3) matrix as the first column of the ($\lceil \sqrt{m_i} \rceil$ $\times$ $\lceil \sqrt{m_i} \rceil$ $\times$ 3) matrix, and then take the ($\lceil \sqrt{m_i} \rceil$ + 1)$^{\rm{th}}$ to 2$\lceil \sqrt{m_i} \rceil ^{\rm{th}}$ elements from top to bottom in the (1 $\times$ $m_i$ $\times$ 3) matrix as the second column of the ($\lceil \sqrt{m_i} \rceil$ $\times$ $\lceil \sqrt{m_i} \rceil$ $\times$ 3) matrix, and so on. Notice that if $m_i$ is not a perfect square number, the last column of the ($\lceil \sqrt{m_i} \rceil$ $\times$ $\lceil \sqrt{m_i} \rceil$ $\times$ 3) matrix cannot be filled. In such a scenario, the few remaining elements will be set to the number zero. After the reshaping process is completed, a ($\lceil \sqrt{m_i} \rceil$ $\times$ $\lceil \sqrt{m_i} \rceil$ $\times$ 3) matrix can be obtained, as shown in \textbf{the middle sub-figure of Figure~\ref{fig:PMSGeneration}}.

The reshaped matrix can be regarded as three ($\lceil \sqrt{m_i} \rceil$ $\times$ $\lceil \sqrt{m_i} \rceil$ $\times$ 1) sub-matrices: [:, :, 0], [:, :, 1], and [:, :, 2], in which the first sub-matrix is actually $V^i_{name}$, and the second as well as the third sub-matrices are $V^i_{value}$ and $V^i_{depth}$, respectively. Among them, the elements of $V^i_{name}$ and $V^i_{value}$ are in the form of characters, and the elements of $V^i_{depth}$ are integers. To make them have a uniform format, we employ an existing method in the reference~\cite{li2020variable} to convert elements of $V^i_{name}$ and $V^i_{value}$ from the character form to the numeric form. Specifically, for a variable's name or a variable's value in the string form (denoted as $str$), we use the following formula to convert it to an integer (denoted as $str_{asc}$):

\begin{equation}
\label{equ:ascii}
str_{asc}=\sum_{i=0}^{\left| str \right|} (i + 1) * Ascii(str[i]),
\end{equation}
where $Ascii()$ gets the Ascii code given a character \syrevisetwo{(the strategy of converting strings to values by employing ascii codes has been adopted by many previous works~\cite{naz2020ascii, mushtaq2019new, vijayan2016ascii}).} The multiplication by ($i$ + 1) is to handle the case where two strings have the same batch of characters but in different order. For example, a string ``$Ee01$'' can be converted to an integer 611, which is calculated by $1 * Ascii(\text{`}E\text{'}) + 2 * Ascii(\text{`}e\text{'}) + 3 * Ascii(\text{`}0\text{'}) + 4 * Ascii(\text{`}1\text{'})$.

Now, each element of the ($\lceil \sqrt{m_i} \rceil$ $\times$ $\lceil \sqrt{m_i} \rceil$ $\times$ 3) matrix is already an integer, \syrevise{this three-channel matrix is ready to be transformed into a colored image.} \syrevise{Here we employ RGB (Red, Green, and Blue), a broadly-used additive color model, to complete this transformation process, because of its prevalence, availability, and simplicity. In RGB, three primary colors each have values ranging from 0 to 255, they can be combined together in various ways to reproduce a broad array of colors in the visible spectrum~\cite{hirsch2005exploring, fairman1997cie}. } \syrevise{To make the ($\lceil \sqrt{m_i} \rceil$ $\times$ $\lceil \sqrt{m_i} \rceil$ $\times$ 3) matrix adapt to the RGB model, we normalize the elements in each of its sub-matrices, namely, [:, :, 0], [:, :, 1], and [:, :, 2] separately, according to the mentioned value range of RGB. The normalization can also alleviate the imbalance among the three dimensions, because the three sub-matrices reflect different resources of memory information thus have different scales. After normalization, the ($\lceil \sqrt{m_i} \rceil$ $\times$ $\lceil \sqrt{m_i} \rceil$ $\times$ 3) matrix can be directly transformed into a colored image, as shown in \textbf{the rightmost sub-figure of Figure~\ref{fig:PMSGeneration}}.}

\syrevisetwo{The motivation for transforming run-time memory information into PMS images is to facilitate the following task of distance calculation. Specifically, memory information is raw data collected for failure representation, its scale could be typically large, and its form is not easy to recognize or handle by machines either. \textbf{Images are a proper candidate form to rephrase raw memory data, because the form of images can effectively integrate various sources of information of memory data into a single block, such a re-representation for raw data can provide well-organized and highly-structured objects for the downstream deep learning technique that performs distance calculation.} Moreover, the form of visualized images can also boost human developers' comprehension of failure indexing results, because images can be smoothly recognized by human beings at a glance.}

\subsection{Model training}
\label{subsect:modeltraining}

The training phase of SURE is illustrated in the upper part of Figure~\ref{fig:overview}. Specifically, for a pair of failed test cases, we first collect the corresponding two sets of  memory information on the faulty program with breakpoints, and then transform them into two PMS images, respectively, as mentioned in Section~\ref{subsect:breakpointdetermination} $\sim$ \ref{subsect:generatePMS}. 

We use the idea of Siamese learning to construct a similarity prediction model for two failed test cases (in the form of PMS images), because Siamese learning is a classical supervised learning strategy in similarity learning tasks~\cite{chicco2021siamese,bromley1993signature,chopra2005learning}. The goal of a Siamese learning-based model is to learn a similarity metric between pairs of input samples, it typically consists of two identical sub-networks for feature extraction that share the same structure and weights, as well as fully connected layers that are responsible for taking the learned representations from the mentioned feature extraction networks, thus producing a final output that reflects the similarity between two input samples. 

The architecture of our Siamese learning model is designed as follows: two identical feature extraction networks (will be specified in the second research question of Section~\ref{subsect:rq}), and two fully connected layers (with a ReLU activation function in between). The loss function we use is the binary cross entropy loss (BCELoss), which is commonly used in deep learning especially for binary classification tasks, as shown in Formula~\ref{equ:bceloss}:
\begin{equation}
	\label{equ:bceloss}
	L=-\frac{1}{N} \sum_{i=1}^N\left[y_i \log \left(p_i\right)+\left(1-y_i\right) \log \left(1-p_i\right)\right],
\end{equation}
where $N$ is the number of the data samples, $y_i$ is the true binary label for the $i\rm{^{th}}$ sample, and $p_i$ is the predicted probability that the $i\rm{^{th}}$ sample is positive. As a reminder, the true binary label of a pair of PMS images is whether ``1'' (i.e., these two failed test cases are triggered by the same fault, thus the similarity between them should be 1) or ``0'' (i.e., these two failed test cases have different root causes, thus the similarity between them should be 0). We denote pairs of PMS images with the label ``1'' as \emph{positive\ samples}, while denote pairs of PMS images with the label ``0'' as \emph{negative\ samples}.

In particular, we use only 30\% of the simulated faults to train the model, and use 70\% of the simulated faults as well as real-world faults to test the model (please refer to Section~\ref{subsect:dataset} for the details of the faults used in the experiments).

\subsection{Prediction}
\label{subsect:prediction}

Once the similarity prediction model is trained, it can be used to predict the similarity between two unseen PMS images (i.e., two failed test cases). This similarity reflects the likelihood of these two failures being triggered by the same fault. Before sending two images to the model, we first preprocess them into a uniform size. This is because handling two images with different sizes is hard for Siamese-based models, in light of the structure of the two feature extraction networks must be identical.

\subsection{Distance calculation}
\label{subsect:distancecalculation}

The output of our Siamese learning model is the similarity between two failures. Before sending the value of the produced similarity to the downstream clustering algorithm, we further process it by Formula~\ref{equ:sizediv}, to make up for the loss caused by harmonizing the sizes of images in the previous step.

\begin{equation}
	\label{equ:sizediv}
	Distance=(1 - Similarity) \times SizeDiv.
\end{equation}

We first explain the first half of this formula, i.e., $1 - Similarity$. The similarity produced by the Siamese model measures how similar two PMS images are. To adapt it for clustering algorithms, we subtract the value of the similarity between two failures from 1, to reflect how dissimilar they are. Next, we explain the multiplication of the factor $SizeDiv$. As mentioned in Section~\ref{subsect:generatePMS}, a PMS image representing the failed test case $f_i$ is a square whose side length is $\lceil \sqrt{m_i} \rceil$, where $m_i$ is the number of variables queried at all breakpoints during the execution of $f_i$. It is obvious that different $f_i$ may correspond to different $m_i$, thus resulting in different side lengths of PMS images. Considering that Siamese models are hard to handle two images with different sizes, in the prediction phase, we preprocess two input images into the same size before feeding them to the model. But as mentioned previously, the side length of PMS images embodies the number of queried variables during running a failed test case, which can reveal the characteristics of a failure to some extent, and thus can also contribute to the distance measurement between two failed test cases. To alleviate this issue, we design $SizeDiv$, a factor that is able to quantitatively reflect the divergence between the sizes of two images. Specifically, $SizeDiv$ is calculated by dividing the side length of the larger image by the side length of the smaller image.

\subsection{Clustering}
\label{subsect:clustering}

In Section~\ref{subsect:failureindexing}, we have pointed out that there is no clustering technique that is universally applicable in uncovering the variety of structures present in multidimensional data sets~\cite{jain1999data}. That is, the clustering algorithm is not the core of failure indexing techniques. Therefore, we employ the clustering component of MSeer~\cite{gao2019mseer}, the state-of-the-art failure indexing technique to date, to complete the clustering phase of SURE. The clustering algorithm is twofold, namely, the faults number estimation and the clustering. The former aims to predict the number of clusters (i.e., the number of faults) given the distance information between data samples (i.e., failed test cases), and the latter is responsible for clustering data samples (i.e., failed test cases) into different groups. Next, we give a concise description of these two steps, for more details please refer to the reference~\cite{gao2019mseer}.

\textbf{For the faults number estimation.} It is well-recognized that one of the trickiest challenges in clustering lies in the estimation of the number of clusters~\cite{fu2020estimating,kingrani2018estimating,tibshirani2001estimating}. Putting it into the context of failure indexing, we can claim that predicting the number of faults given a number of failures is very important. The adopted clustering algorithm presented a novel mountain method-based approach inspired by the previous works~\cite{chiu1994fuzzy,yager1994approximate}, to perform the faults number estimation and the assignment of initial medoids to these clusters at the same time. Specifically, the adopted clustering algorithm first calculates a potential value for each failed test case according to the density of its surroundings, such a potential value is used to measure the possibility of a data point being set as a medoid. And then, 1) The failure with the highest potential value will be chosen as the first medoid. 2) The potential values of all failed test cases will then be updated in accordance with their distance from the newest medoid. 3) Repeating the above two steps iteratively, until the maximum potential value falls within a certain threshold.

\textbf{For the clustering.} Once the number of clusters and the initial medoids are determined, all failures are ready to be clustered. The adopted clustering algorithm utilizes K-medoids, a widely-used clustering strategy, to complete this process. The K-medoids strategy sets actual (not virtual) data points as medoid and thus can be more applicable to SURE, because the mean of memory information is difficult to define. Moreover, the K-medoids strategy has been shown to be very robust to the existence of noise or outliers~\cite{kaufman2009finding}.

\subsection{Running example}
\label{subsect:runningexample}

In the motivating example in Section~\ref{sect:motivatingexample}, we use a toy program to show that the most prevalent and advanced approach to date can still not good enough to produce satisfactory failure indexing results. That toy program successfully reveals the bottleneck of existing approaches with only 17 program statements. But such a small-scale program is not suitable to illustrate the workflow of SURE, because a small number of program statements will correspondingly result in a small number of breakpoints, thus further resulting in a small number of queried variables. In such cases, the generated PMS images will be very small, which does not affect the running of SURE but can hinder readers' comprehension as a running example. Therefore, here we use a complete running example to illustrate the workflow of SURE, where CC and SD do not work.

We employ Grep, a classical tool aiming to print lines in specific files that contain a match of the given patterns. We obtain Grep (v2.4) which contains 13,274 lines from the Software Infrastructure Repository (SIR)~\cite{SIRrepo}, and randomly mutate three positions of the original clean program to inject three faults into the program, to obtain a 3-bug faulty version. The three mutants are shown in Listing~\ref{list:exampleMutants}.

\begin{lstlisting}[language=C, escapeinside={(*@}{@*)}]
 3164	  if (leftoversf)
 3165		{
 3166			copyset(leftovers, labels[ngrps]);
 3167			copyset(intersect, labels[j]);
(*@\colorbox{deleteColor}{
3168 -\qquad\quad\ \,\,MALLOC(grps[ngrps].elems, position, d->nleaves);\qquad\qquad\qquad\qquad\qquad\qquad\qquad\qquad\qquad\qquad\quad
}@*)
(*@\colorbox{addColor}{
\qquad\ +\qquad\quad\ \,\,MALLOC(grps[ngrps].elems, position, d->nleaves*-1);\qquad\qquad\qquad\qquad\qquad\qquad\qquad\qquad\qquad\ \ \ 
}@*)
 3169			copy(&grps[j], &grps[ngrps]);
 3170			++ngrps;
 3171	     }
          ...
 4681     for (ep = text + size - 11 * len;;)
 4682    	{
 4683    		while (tp <= ep)
 4684    			{
(*@\colorbox{deleteColor}{
 4685 -\qquad\qquad\qquad\ \ \ \,\,d = d1[U(tp[-1])], tp += d;\qquad\qquad\qquad\qquad\qquad\qquad\qquad\qquad\qquad\qquad\qquad\qquad\qquad\quad\ 
}@*)
(*@\colorbox{addColor}{
\qquad\ + \qquad\qquad\qquad\ \,\,  			   d = d1[U (! tp[-1])], tp += d;\qquad\qquad\qquad\qquad\qquad\qquad\qquad\qquad\qquad\qquad\qquad\qquad\qquad
}@*)
 4686					 d = d1[U(tp[-1])], tp += d;
 4687				     if (d == 0)
 4688			  			goto found;
           ...
 7480	   compile_stack.stack = TALLOC (INIT_COMPILE_STACK_SIZE, compile_stack_elt_t);
 7481	   if (compile_stack.stack == NULL)
 7482 	      return REG_ESPACE;
(*@\colorbox{deleteColor}{
 7484 -\quad\,\,\,compile\_stack.size = INIT\_COMPILE\_STACK\_SIZE;\qquad\qquad\qquad\qquad\qquad\qquad\qquad\qquad\qquad\qquad\qquad\qquad\quad
}@*)
(*@\colorbox{addColor}{
    \qquad\ +\quad\,\,\,compile\_stack.size = 0;\qquad\qquad\qquad\qquad\qquad\qquad\qquad\qquad\qquad\qquad\qquad\qquad\qquad\qquad\qquad\qquad\qquad\qquad
}@*)
 7485	   compile_stack.avail = 0;

\end{lstlisting}

 We compile the mutated program and run the accompanying test suite, observing a series of failures and also getting a number of passed test cases. Then, we conduct failure indexing following the aforementioned steps: \syrevise{For the breakpoint determination (Section~\ref{subsect:breakpointdetermination}), we employ an SBFL technique (e.g., DStar~\cite{wong2013dstar}) to determine the suspiciousness of program statements and setting breakpoints. For the collection of run-time program memory (Section~\ref{subsect:collectionmemory}), we run failed test cases and collect the run-time memory information $MI$ in accordance with Formula~\ref{equ:mii}. $MI$ comprises variables' names, variables' values, and the depth of the stack frame, whose content and structure are defined as Formula~\ref{equ:vnameij} $\sim$ Formula~\ref{equ:vdepthi}, and $MI$'s scale is determined by Formula~\ref{equ:mi}. For the generation of PMS images (Section~\ref{subsect:generatePMS}), we first convert variables' names and values to the numeric form using Formula~\ref{equ:ascii}, and then transform each set of memory information into a PMS image.}
 	
\begin{figure}[t]
	\centering
	\includegraphics[width=\linewidth]{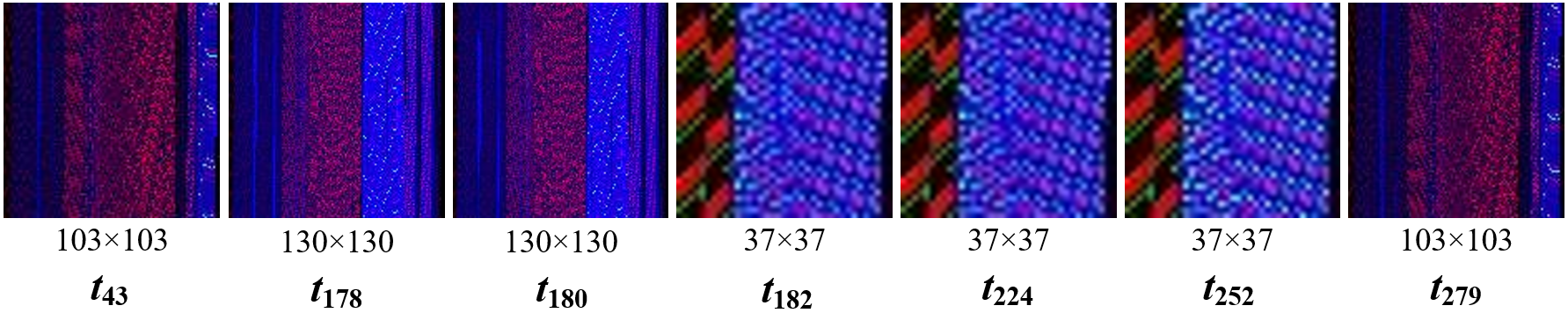}
	\caption{The PMS images of the failed test cases}
	\label{fig:images_runningExample}
\end{figure}

\syrevise{It is hard and not necessary to completely illustrate this running example because there are too many failed test cases. Here we randomly select seven of them, namely, $t_{43}$, $t_{178}$, $t_{180}$, $t_{182}$, $t_{224}$, $t_{252}$, and $t_{279}$, to show their PMS images and describe the following steps\footnote{The remaining failures omitted here can also support the conclusion of this example.}. Among them, $t_{43}$ and $t_{279}$ are triggered by the mutant located in Line 3168, $t_{178}$ and $t_{180}$ are triggered by the mutant located in Line 4685, and $t_{182}$, $t_{224}$ as well as $t_{252}$ are triggered by the mutant located in Line 7484. The seven PMS images as well as their size, are given in Figure~\ref{fig:images_runningExample}. Then, we employ the trained model (Section~\ref{subsect:modeltraining}) to predict the similarity between pairs of failures (Section~\ref{subsect:prediction}), and get the distance information (Section~\ref{subsect:distancecalculation}), as given in Table~\ref{tab:distanceExample}. Finally, all failures are clustered into different groups using the two phases of the clustering algorithm: the faults number estimation and the clustering (Section~\ref{subsect:clustering}). The outcome of the clustering algorithm is: \{$t_{43}$, $t_{279}$\}, \{$t_{178}$, $t_{180}$\}, and \{$t_{182}$, $t_{224}$, $t_{252}$\}, which is consistent with the characteristic of the distance information given in Table~\ref{tab:distanceExample}. Obviously, this clustering outcome achieves promising failure indexing since the number of clusters (i.e., the number of faults, three) is correctly estimated, and all failures are grouped correctly according to the root cause.}

Moreover, we can observe that the characteristics of the PMS images in Figure~\ref{fig:images_runningExample} are highly identical to the failure indexing result. Specifically, the PMS images of $t_{43}$ and $t_{279}$ are similar and have the same size. Likewise, we can observe the same phenomenon on \{$t_{178}$, $t_{180}$\} and \{$t_{182}$, $t_{224}$, $t_{252}$\}. With the support of the PMS images, developers can be easily convinced by the failure indexing outcome and thus adopt it.

\begin{table}
	\caption{\syrevisetwo{Distance information between failures}}
	\label{tab:distanceExample}
	\begin{tabular}{p{0.7cm}p{0.7cm}p{0.7cm}p{0.7cm}p{0.7cm}p{0.7cm}p{0.7cm}p{0.7cm}}
		\toprule
		& \bm{$t_{43}$} & \bm{$t_{178}$} & \bm{$t_{180}$} & \bm{$t_{182}$} & \bm{$t_{224}$} & \bm{$t_{252}$} & \bm{$t_{279}$} \\
		\midrule
		\bm{$t_{43}$} & 0.00 & 1.26 & 1.26 & 2.78 & 2.78 & 2.78 & \cellcolor{gray!50}0.01\\
		\bm{$t_{178}$} & 1.26 & 0.00 & \cellcolor{gray!50}0.02 & 3.51 & 3.51 & 3.51 & 1.26 \\
		\bm{$t_{180}$} & 1.26 & \cellcolor{gray!50}0.02 & 0.00 & 3.51 & 3.51 & 3.51 & 1.26 \\
		\bm{$t_{182}$} & 2.78 & 3.51 & 3.51 & 0.00 & \cellcolor{gray!50}0.01 & \cellcolor{gray!50}0.02 & 2.78  \\
		\bm{$t_{224}$} & 2.78 & 3.51 & 3.51 & \cellcolor{gray!50}0.01 & 0.00 &  \cellcolor{gray!50}0.02 & 2.78 \\
		\bm{$t_{252}$} & 2.78 & 3.51 & 3.51 & \cellcolor{gray!50}0.02 & \cellcolor{gray!50}0.02 & 0.00 & 2.78 \\
		\bm{$t_{279}$} & \cellcolor{gray!50}0.01 & 1.26 & 1.26 & 2.78 & 2.78 & 2.78 & 0.00 \\
		\bottomrule
	\end{tabular}
\end{table}

\section{Experimental Setup}
\label{sect:setup}

\subsection{Research Question}
\label{subsect:rq}

\hangafter 1
\hangindent 3.4em
$\bullet$ \syrevise{\textbf{RQ1: Does the value of the breakpoint determination threshold impact SURE's effectiveness?}}

As mentioned in Section~\ref{subsect:breakpointdetermination}, SURE will take the Top-$x$\% riskiest program statements as breakpoints. We investigate how the value of $x$ impacts the effectiveness of SURE. Specifically, we compare the effectiveness of SURE when such a threshold is set from 0\% to 100\% with 10\% increments, i.e., 10\%, 20\%, 30\%, ..., 100\%.

\vspace{3pt}
\hangafter 1
\hangindent 4.4em
$\bullet$ \syrevise{\textbf{RQ2: Does the selection of the feature extraction network impact SURE's effectiveness?}}

For a pair of failures, SURE uses a Siamese learning-based model to predict its distance. Because the input samples in our context are in the form of PMS images, the feature extraction network is correspondingly convolutional. In this RQ, we investigate how the feature extraction network impacts the effectiveness of SURE. Specifically, we compare the effectiveness of SURE when the network configures VGG-16~\cite{simonyan2014very}, AlexNet~\cite{krizhevsky2012imagenet}, and ResNet-18~\cite{he2016deep}.

\vspace{3pt}
\hangafter 1
\hangindent 4.3em
$\bullet$ \syrevise{\textbf{RQ3: How does SURE perform compared with the most prevalent and advanced failure indexing technique?}}

 In Section~\ref{sect:background}, we have introduced that the CC-based strategy has been extensively used by the failure indexing community due to its simplicity, and the SD-based strategy has been recognized as the state-of-the-art solution. Thus, we compare SURE with these two types of techniques for more convincing evaluation.

For the CC-based strategy, we select \bm{$Cov_{hit}$} as our baseline, because it is the most common technique in the CC class. Specifically, it represents a failed test case as a binary numeric vector with a length equal to the number of program statements, the $i^{\rm{th}}$ element of the vector will be set to 1 if the $i^{\rm{th}}$ statement is covered by this failed test case, and 0 otherwise. Moreover, seeing that there have emerged some works concerning the impact of the execution frequency of program statements on debugging~\cite{shu2016fault,wen2012software,vancsics2023new}, we also adopt another technique in the CC class, \bm{$Cov_{count}$}, as our baseline. Specifically, $Cov_{count}$ also represents a failed test case as a numeric vector with a length equal to the number of program statements, while the $i^{\rm{th}}$ element of the vector will be set to the actual execution frequency if the $i^{\rm{th}}$ statement is covered (rather than a constant, 1), and 0 otherwise.
	
For the SD-based strategy, we select \bm{$MSeer$} as our baseline, because it is the state-of-the-art technique not only in the SD class but also in the current field of failure indexing. Specifically, it first runs a failed test case along with passed test cases against the faulty program, and inputs the gathered coverage information to Crosstab~\cite{wong2012towards}, an SBFL technique, to calculate the suspiciousness of being faulty for each program statement. And then, ranking all statements in descending order according to their suspiciousness values. Such a ranking list will serve as the proxy for this failed test case. Moreover, in our previous empirical study, we comprehensively investigated the factors that could impact the effectiveness of MSeer, finding that the selection of SBFL formulas matters~\cite{song2022comprehensive}. According to our result, if Crosstab is replaced by GP19 (another SBFL formula evolved by genetic programming in the reference~\cite{yoo2012evolving}), MSeer will do much better. Thus, we manually upgrade MSeer to \bm{$MSeer_{GP19}$}, which employs GP19 to calculate suspiciousness, and include it as our baseline\footnote{As a reminder, $MSeer_{GP19}$ is not an existing published failure indexing technique, it is manually created by us to further evaluate the competitiveness of SURE.}.

To summarize, we compare SURE with four techniques, i.e., $Cov_{hit}$ and $Cov_{count}$ (the CC-based strategy), as well as $MSeer$ and $MSeer_{GP19}$ (the SD-based strategy).

\vspace{3pt}
\hangafter 1
\hangindent 4.3em
$\bullet$ \syrevise{\textbf{RQ4: To what extent can SURE help human developers comprehend failure indexing results?}}

As a visualized failure indexing technique, SURE represents failed test cases as human-friendly PMS images. In this way, developers who receive the failure indexing result can not only know how all failures are grouped, but also be aware of why they are grouped in the current manner. In this RQ, we quantitatively investigate to what extent developers can \syrevise{comprehend} the failure indexing result with the support of PMS images. Similar to RQ3, we compare the \syrevise{comprehensibility} of SURE with that of CC-based and SD-based techniques. Specifically, we conduct a comparison of developers' performance on \syrevise{comprehending} failure indexing results among given PMS images (i.e., the fingerprinting function of SURE), code coverage vectors (i.e., the fingerprinting function of CC-based strategies), and suspiciousness ranking lists (i.e., the fingerprinting function of SD-based strategies). This human study involves 9 tasks and 15 experienced graduate students in Computer Science from \syrevise{Wuhan University}.

\subsection{Parameter Setting}
\label{subsect:parameter}

SURE needs to first determine the suspiciousness value of program statements, and thus sets breakpoints at those highly-risky positions. In this phase, we use DStar~\cite{wong2013dstar}, the state-of-the-art SBFL technique that has been utilized broadly. Considering the preference for DStar in many other studies (such as the references~\cite{arrieta2018spectrum,pearson2017evaluating,widyasari2022real}), we set the value of * in DStar to 2, the most thoroughly-explored value, in our experiments. Such a choice is not hard-coded but can be configurable, any other fault localization techniques working at the granularity of statements can be adapted to this phase.

In the training phase, we set the value of batch size as 16. The initial value of the learning rate is defined as 1e-4, and it will be multiplied by 0.96 after each epoch.

\subsection{Dataset}
\label{subsect:dataset}

For comprehensive and robust evaluation, in the experiments, we use the benchmark involving both simulated faults and real-world faults. For simulated benchmarks, we manually seed different types of faults into clean programs, to obtain simulated faulty programs, in light of the fact that previous research has confirmed that mutation-based faults can provide credible results for experiments in software testing and debugging~\cite{andrews2005mutation,andrews2006using,do2006use,just2014mutants,liu2006statistical,pradel2018deepbugs}. The clean programs are obtained from the Software Infrastructure Repository (SIR)~\cite{do2005supporting}. For real-world benchmarks, we search for the Defects4J projects~\cite{just2014defects4j}, one of the most popular datasets in the current field of software testing and debugging, according to the test case transplantation strategy proposed by An et al.~\cite{an2021searching}. All faulty programs contain one, two, three, four, or five bugs (also referred to as 1-bug, 2-bug, 3-bug, 4-bug, and 5-bug faulty versions, respectively), because such numbers of faults are most-widely investigated in previous studies~\cite{gao2019mseer,le2015should,xiaobo2021test,zakari2019parallel}.

\subsubsection{Simulated faulty programs}
\label{subsubsect:SIR}

SIR is a classical repository in the community of software testing and debugging, which has been employed in numerous pioneering studies~\cite{yoo2010using,wong2011effective,zakari2020multiple,huang2022dissimilarity,kim2022learning}. From SIR we obtain Flex, Grep, Gzip, and Sed, \syrevise{four projects that have been extensively adopted in earlier works~\cite{bertolino2017adaptive, ghandehari2018combinatorial, song2022evolving}}, as the clean programs to mutate (i.e., seed faults). The concise information about them can be found in Table~\ref{tab:benchmark}. We utilize an open-source tool with ``13'' fork and ``23'' star on GitHub to perform mutation, which defines 67 types of points that can be mutated, and provides several mutation operators for each~\cite{mutationtool}. The mutation operators we leverage can be categorized into the following two classes:

\begin{itemize}
	\item \textbf{\emph{Assignment Fault}}~\cite{jeffrey2008fault}: Editing the value of constants in the statement, or replacing the operators such as addition, subtraction, multiplication, division, etc. with each other.
	
	\item \textbf{\emph{Predicate Fault}}~\cite{xuan2016nopol}: Reversing the $if$-$else$ predicate, or deleting the $else$ statement, or modifying the decision condition, and so on. 
\end{itemize}

To create an $r$-bug faulty version ($r$ = 2, 3, 4, 5), the faults from $r$ individual 1-bug faulty versions are injected into the same program. Such a strategy has been adopted \syrevise{by the majority of the} published studies in the field of multi-fault debugging~\cite{abreu2009spectrum,hogerle2014more,huang2013empirical,lamraoui2016formula,yu2015does}. In total, we obtain 1,000 SIR faulty versions containing 1$\sim$5 faults.

\begin{table}[]
	\centering
	\caption{\label{tab:benchmark} Benchmarks}
	\begin{tabular}{p{1.6cm}p{1.5cm}p{1.5cm}p{1.5cm}p{3.5cm}}
		\hline
		\rule{0pt}{6pt} \textbf{Language} &  \textbf{Project} &  \textbf{Version} &  \textbf{kLOC}   &  \textbf{Functionality}  \\ \hline
		\multirow{4}{*}{C}        &   Flex             & 2.5.3            & 14.5            & Parser generator \\
		&   Grep             & 2.4              & 13.5           & Text matcher \\
		&   Gzip             & 1.2.2            & 7.3             & File archiver \\
		&   Sed              & 3.02             & 10.2           & Stream editor \\
		
		\rule{0pt}{9pt}\multirow{5}{*}{Java}   &   Chart           & 2.0.0            & 96.3          & Chart library   \\
		&   Closure        & 2.0.0            & 90.2          & Closure compiler  \\
		&   Lang            & 2.0.0            & 22.1            & Apache commons-lang  \\
		&   Math            & 2.0.0            & 85.5           & Apache commons-math \\
		&   Time            & 2.0.0            & 28.4           & Date and time library \\
		\hline	
	\end{tabular}
\end{table}

\subsubsection{Real-world faulty programs}
\label{subsubsect:D4J}

Defects4J is one of the most popular benchmarks in the current field of software testing and debugging, due to its realism and ease of use~\cite{just2014defects4j}. Defects4J is typically for single-fault scenarios, namely, no matter how many bugs are contained in a faulty program, the provided test suite is only sufficient to reveal one of them. Such a characteristic hinders its use in failure indexing studies. To adapt Defects4J to multi-fault scenarios, An et al. presented a test case transplantation strategy~\cite{an2021searching}. Specifically, the majority of Defects4J faulty versions are indexed chronologically according to the revision date, a lower ID indicates a more recent version. Therefore, the fault in a newer version is also likely to be contained in an older version. For example, the fault of the faulty version Math-5b is found to exist in the faulty version Math-6b as well, it is not revealed in Math-6b simply due to the absence of the fault-revealing test case. If this test case is transplanted to Math-6b, the enhanced test suite is able to reveal both of the two faults, thus a 2-bug faulty version can be obtained~\cite{an2021searching}. Following this strategy, we search for a collection of multi-fault Defects4J programs, from Chart, Closure, Lang, Math, and Time, as shown in Table~\ref{tab:benchmark}. \syrevise{Because Defects4J multi-fault versions are obtained by searching in real-life environments, their number is not very large.} In total, we obtain 100 Defects4J faulty versions containing 1$\sim$5 faults.

\subsection{Metric}
\label{subsect:metric}

The mission of a failure indexing technique is to identify the mutual relationship among failures by clustering, i.e., determine which failures are triggered by the same (or different) fault(s). Correspondingly, the outcome of failure indexing is several clusters of failed test cases. For evaluation, we need to quantitatively measure to which extent the delivered clusters correctly reflect the true relationship among failures. There are two typical types of metrics in evaluating clustering outcomes, namely, external metrics~\cite{wu2009adapting} and internal metrics~\cite{tan2016introduction}. The former compares clustering results with the oracle (true linkages between failures and faults), while the latter examines inherent properties of the delivered clusters, such as compactness and separation, etc., typically when the oracle is inaccessible. In our controlled experiments, the oracle clusters can be obtained easily in advance. Specifically, for an $r$-bug SIR faulty version, it is generated by combining $r$ individual single-bug faulty versions. We can run the test suite against these $r$ faulty versions separately, thus becoming aware of the culprit fault of each failure. And for an $r$-bug Defects4J faulty version, it is generated by transplanting the failed test cases triggered by a fault to the faulty version that contains another fault. This transplantation process itself explicitly indicates the relationship between failures and underlying faults. Thus, we use external metrics to measure the effectiveness of  failure indexing techniques. Concretely speaking, we employ four external metrics, the Fowlkes and Mallows Index (FMI), the Jaccard Coefficient (JC), the Precision Rate (PR), and the Recall Rate (RR), in our experiments, because they are all classical metrics for clustering and have been adopted by many published studies~\cite{wu2009adapting,xie2017new,zhang2022smart}. Among them, FMI and JC are pair of test cases-based, while PR and RR are single test case-based.

\subsubsection{Pair of test cases-based metrics}
\label{subsubsect:fmijc}

The pair of test cases-based metric refers to comparing the indexing consistency of each pair of failed test cases in the generated cluster with the oracle cluster. Four possible scenarios in the comparison are given in Table~\ref{tab:pairmetric}. Specifically, supposing that there are $n$ failed test cases that need to be clustered, they can form $C^2_n$ pairs by combining any two ones. For a pair, if its two member failures are determined to be triggered by the $Same$ fault in the generated cluster, and these two are also divided into the $Same$ group in the oracle cluster, this pair falls into the category of ``$SS$''. Similarly, a pair is also possible to fall into the categories of ``$SD$'', ``$DS$'', or ``$DD$''. We use $X_{SS}$, $X_{SD}$, $X_{DS}$, and $X_{DD}$ to denote the numbers of the pairs fallen into  ``$SS$'', ``$SD$'', ``$DS$'', and``$DD$'', respectively. Obviously, the sum of $X_{SS}$, $X_{SD}$, $X_{DS}$, and $X_{DD}$ will be $C^2_n$, because any pair of failures will belong to one of the four scenarios. FMI and JC can integrate these four notations into one metric, to comprehensively reflect the similarity between the generated cluster and the oracle cluster, as shown in Formula~\ref{equ:fmi} and Formula~\ref{equ:jc}, respectively.

\begin{equation}
	\label{equ:fmi}
	FMI = \sqrt{\frac{X_{SS}}{X_{SS} + X_{SD}} \times \frac{X_{SS}}{X_{SS} + X_{DS}}},
\end{equation}
\vspace{6pt}
\begin{equation}
	\label{equ:jc}
	JC = \frac{X_{SS}}{X_{SS} + X_{SD} + X_{DS}}.
\end{equation}

\vspace{3pt}
It can be proved that the intervals of FMI and JC are both [0, 1], and that the larger the value in this range, the more effective clustering is.

\subsubsection{Single test case-based metrics}
\label{subsubsect:prrr}

\begin{table}[]
	\centering
	\setlength{\belowcaptionskip}{3pt}
	\caption{\label{tab:pairmetric} Four scenarios in the pair of test cases-based metrics} 
	\begin{tabular}{cll}
		\hline
		\multirow{2}{*}{\textbf{Notation}} & \multicolumn{2}{c}{\textbf{Results of failure indexing}}                           \\ \cline{2-3}  & 
		In the generated cluster & In the oracle cluster \\ \hline
		SS                       & Same                            & Same                         \\
		SD                      & Same                            & Difference              \\
		DS                      & Difference                  & Same                         \\
		DD                      & Difference                 & Difference       \\ \hline         
	\end{tabular}
\end{table}

\begin{table}[]
	\centering
	\setlength{\belowcaptionskip}{3pt}
	\caption{\label{tab:singlemetric} Four scenarios in the single test case-based metrics} 
	\begin{tabular}{cll}
		\hline
		\multirow{2}{*}{\textbf{Notation}} & \multicolumn{2}{c}{\textbf{Results of failure indexing}}                           \\ \cline{2-3}  & 
		In the generated cluster & In the oracle cluster \\ \hline
		TP                       & Positive                     & Positive                         \\
		FP                       & Positive                     & Negative               \\
		TN                      & Negative                   & Negative                         \\
		FN                      &Negative                    & Positive        \\ \hline         
	\end{tabular}
\end{table}

The single test case-based metric refers to comparing the classification result of each failed test case in the generated cluster with the oracle cluster. Four possible scenarios in the comparison are given in Table~\ref{tab:singlemetric}. Specifically, for a failed test case, if it is predicted to be triggered by a fault in the generated cluster (i.e., $Positive$), and its root cause is indeed this fault (i.e., $True$ prediction result), this failure falls into the category of ``$TP$''. Likewise, if a failed test case is predicted not to be triggered by a fault in the generated cluster (i.e., $Negative$), while its root cause is actually this fault (i.e., $False$ prediction result), this failure falls into the category of ``$FN$''.  Similarly, a failed test case is also possible to fall into the categories of ``$FP$'' and ``$TN$''. We use $X_{TP}$, $X_{FN}$, $X_{FP}$, and $X_{TN}$ to denote the numbers of the failed test cases fallen into ``$TP$'', ``$FN$'', ``$FP$'', and ``$TN$'', respectively. Obviously, the sum of ``$TP$'', ``$FN$'', ``$FP$'', and ``$TN$'' will be the number of failures, because any failure will belong to one of the four scenarios. PR and RR can integrate these four notations into one metric, to comprehensively reflect the similarity between the generated cluster and the oracle cluster, as shown in Formula~\ref{equ:pr} and Formula~\ref{equ:rr}, respectively.

\begin{equation}
	\label{equ:pr}
	PR = \frac{X_{TP}}{X_{TP} + X_{FP}},
\end{equation}
\vspace{6pt}
\begin{equation}
	\label{equ:rr}
	RR = \frac{X_{TP}}{X_{TP} + X_{FN}}.
\end{equation}

\vspace{3pt}
It can be proved that the intervals of PR and RR are both [0, 1], and that the larger the value in this range, the more effective clustering is.

Notice that FMI, JC, PR, and RR evaluate the effectiveness of clustering of one faulty version. In our experiments, there are a series of faulty versions to make our evaluation abundant. Thus, for a failure indexing technique $T$,  we determine its effectiveness by adapting the FMI, JC, PR, and RR metrics on one faulty version to that on multiple faulty versions:

\begin{equation}
	S\_ {M}_M^T=\sum_i^{V_{e q u a l}^T} M_i,
\end{equation}
where ``$S\_M$'' is the abbreviation for ``$Sum\_Metrics$''. $V^T_{equal}$ is the number of faulty versions whose number of faults is correctly predicted by using $T$. $M_i$ is the metric value ($M$ takes FMI, JC, PR, or RR) on the $i^{\rm{th}}$ ``$k$ == $r$'' faulty version\footnote{For an $r$-bug faulty version, if the predicted number of faults $k$ is equal to $r$, we label this faulty version as ``$k$ == $r$''.}. As a reminder, an $r$-bug faulty version contains $r$ faults, which is hard to know in advance. A failure indexing technique should first correctly predict the number of faults $r$, and then cluster all failures into $r$ groups. For an $r$-bug faulty version, if the predicted number of faults $k$ is not equal to $r$, we do not evaluate the clustering effectiveness on this faulty version, because prior studies have pointed out that the performance of a failure indexing technique can be mainly determined by those ``$k$ == $r$'' faulty versions~\cite{gao2019mseer,song2022comprehensive}, that is, the contribution of ``$k$ != $r$'' ones is marginal. And also, it is indeed difficult to compare $k$ delivered clusters with $r$ oracle clusters. Notice that ``$k$ == $r$'' is just an ideal scenario (not necessary) for SURE. Even if $k$ != $r$, SURE can also work. We will further explain this point in Section~\ref{sect:threats}.

In summary, there are five evaluation metrics used in the experiments, which correspond to the two goals of failure indexing mentioned previously. Specifically, $V^T_{equal}$ reflects the goal of \emph{correct faults number estimation}, and $S\_M^T_{FMI}$, $S\_M^T_{JC}$, $S\_M^T_{PR}$, and $S\_M^T_{RR}$ reflect the goal of \emph{promising clustering}.

\subsection{Environment}
\label{subsect:environment}

We collect program coverage and run-time memory information on Ubuntu 16.04.1 LTS with GCC 5.4.0 and JDB 1.8. The Siamese neural network model is trained and deployed on 4 GPUs of GeForce RTX 2080 Ti. The clustering process runs on a server equipped with 96 Intel Xeon(R) Gold 5218 CPU cores with 2.30GHz and 160 GB of memory.

\begin{table}[t]
	\centering
	\renewcommand\arraystretch{1.17}
	\setlength{\belowcaptionskip}{3pt}
	\caption{\label{tab:result_rq1_sir} Performance of SURE on SIR} 
	\resizebox{0.9\textwidth}{!}{
		\begin{tabular}{|cc|c|c|c|c|c|c|c|c|c|c|}
			\hline
			\multicolumn{2}{|c|}{{\cellcolor{gray!20}}}                                  & \cellcolor{gray!20}\textbf{10\%}   & \cellcolor{gray!20}\textbf{20\%}   & \cellcolor{gray!20}\textbf{30\%}   & \cellcolor{gray!20}\textbf{40\%}   & \cellcolor{gray!20}\textbf{50\%}   & \cellcolor{gray!20}\textbf{60\%}   & \cellcolor{gray!20}\textbf{70\%}   & \cellcolor{gray!20}\textbf{80\%}   & \cellcolor{gray!20}\textbf{90\%}   & \cellcolor{gray!20}\textbf{100\%}  \\ \hline
			
			\multicolumn{1}{|c|}{\cellcolor{gray!20}}   & $\bm{V^T_{equal}}$ & \textbf{161}    & 147    & 122    & 143    & 153    & 164    & 142    & 145    & 151    & 155    \\ \cline{2-12} 
			\multicolumn{1}{|c|}{\cellcolor{gray!20}}                         & $\bm{{S\_M}^T_{FMI}}$   & \textbf{126.78} & 114.32 & 96.11  & 114.36 & 121.28 & 131.10  & 112.59 & 112.26 & 118.91 & 122.75 \\ \cline{2-12} 
			\multicolumn{1}{|c|}{\cellcolor{gray!20}}                         & $\bm{{S\_M}^T_{JC}}$     & \textbf{106.39} & 94.79  & 80.36  & 96.97  & 102.25 & 111.44 & 94.95  & 93.18  & 99.82  & 103.33 \\ \cline{2-12} 
			\multicolumn{1}{|c|}{\cellcolor{gray!20}}                         & $\bm{{S\_M}^T_{PR}}$     & \textbf{120.24} & 106.67 & 91.60   & 111.40  & 115.48 & 127.75 & 108.06 & 110.88 & 119.31 & 123.99 \\ \cline{2-12} 
			\multicolumn{1}{|c|}{\multirow{-5}{*}{\cellcolor{gray!20}\textbf{VGG-16}}}                         & $\bm{{S\_M}^T_{RR}}$     & \textbf{80.67}  & 71.20   & 59.52  & 72.80   & 75.92  & 84.55  & 69.20   & 63.20   & 72.66  & 73.17  \\ \hline

			\multicolumn{1}{|c|}{\cellcolor{gray!20}} & \bm{$V^T_{equal}$} & \textbf{167}   & 143    & 139    & 169    & 187    & 160    & 160    & 176    & 163    & 150    \\ \cline{2-12} 
			\multicolumn{1}{|c|}{\cellcolor{gray!20}}  & $\bm{{S\_M}^T_{FMI}}$    & \textbf{131.15} & 113.31 & 109.04 & 135.30  & 148.08 & 125.76 & 126.53 & 137.60  & 127.59 & 118.99 \\ \cline{2-12} 
			\multicolumn{1}{|c|}{\cellcolor{gray!20}}                          & $\bm{{S\_M}^T_{JC}}$     & \textbf{109.70}  & 95.43  & 91.11  & 115.11 & 124.54 & 105.42 & 106.46 & 115.04 & 106.70  & 100.56 \\ \cline{2-12} 
			\multicolumn{1}{|c|}{\cellcolor{gray!20}}                          & $\bm{{S\_M}^T_{PR}}$     & \textbf{123.10}  & 109.06 & 102.31 & 131.53 & 145.66 & 124.69 & 127.87 & 136.72 & 127.4  & 120.04 \\ \cline{2-12} 
			\multicolumn{1}{|c|}{\multirow{-5}{*}{\cellcolor{gray!20}\textbf{AlexNet}}}                         & $\bm{{S\_M}^T_{RR}}$    & \textbf{80.44}  & 70.11  & 69.80   & 86.33  & 89.91  & 76.78  & 80.89  & 82.49  & 79.07  & 74.08  \\ \hline

			\multicolumn{1}{|c|}{\cellcolor{gray!20}}    & $\bm{V^T_{equal}}$& \textbf{118}    & 153    & 151    & 157    & 150    & 161    & 157    & 147    & 172    & 137    \\ \cline{2-12} 
			\multicolumn{1}{|c|}{\cellcolor{gray!20}}                           & $\bm{{S\_M}^T_{FMI}}$   & \textbf{92.71}  & 121.25 & 119.68 & 125.56 & 115.85 & 126.42 & 122.81 & 115.38 & 136.39 & 109.13 \\ \cline{2-12} 
			\multicolumn{1}{|c|}{\cellcolor{gray!20}}                           & $\bm{{S\_M}^T_{JC}}$     & \textbf{77.48}  & 102.14 & 100.77 & 106.47 & 95.82  & 106.07 & 102.51 & 96.42  & 114.98 & 92.45  \\ \cline{2-12} 
			\multicolumn{1}{|c|}{\cellcolor{gray!20}}                         & $\bm{{S\_M}^T_{PR}}$     & \textbf{82.63}  & 115.51 & 115.36 & 122.86 & 114.62 & 121.60  & 121.79 & 116.69 & 135.98 & 109.16 \\ \cline{2-12} 
			\multicolumn{1}{|c|}{\multirow{-5}{*}{\cellcolor{gray!20}\textbf{ResNet-18}}}                        & ${\bm{S\_M}^T_{RR}}$     & \textbf{59.37}  & 79.27  & 76.55  & 78.96  & 68.59  & 77.27  & 73.57  & 68.52  & 84.51  & 67.48  \\ \hline
			
	\end{tabular}}
\end{table}

\begin{figure}[t]
	\centering
	\includegraphics[width=0.9\linewidth]{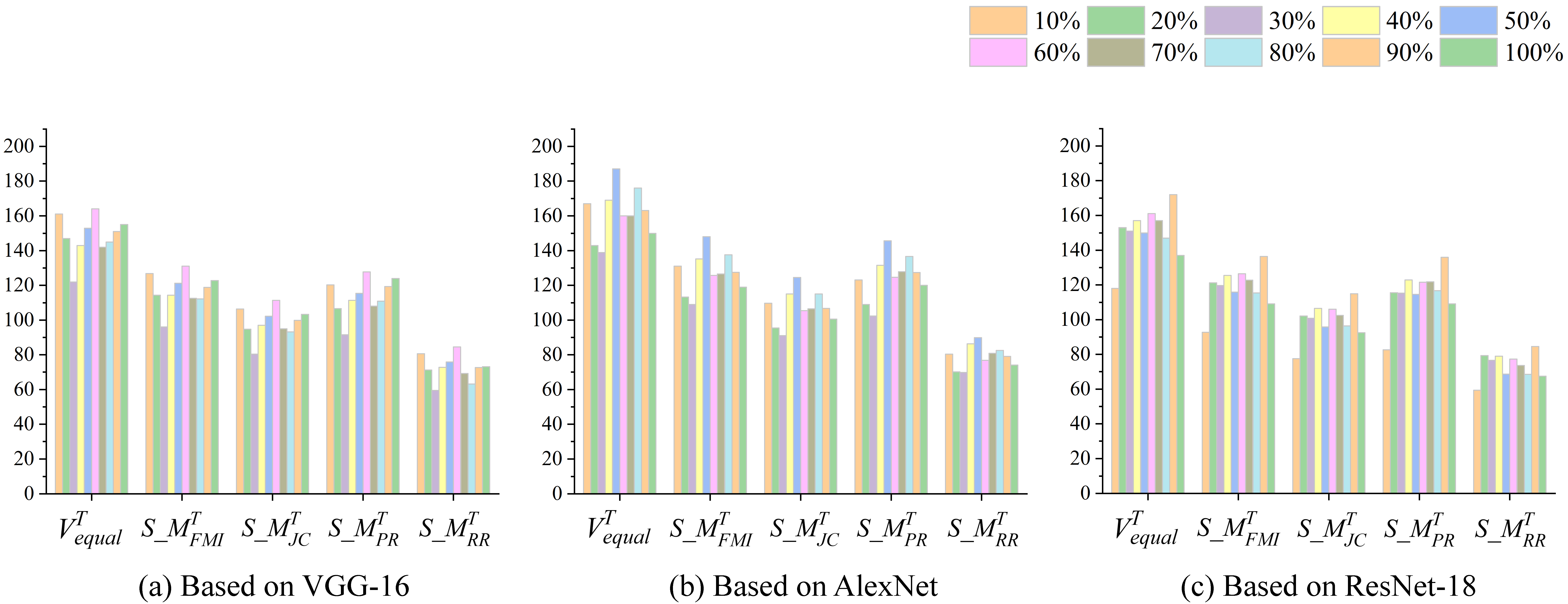}
	\caption{The impact of the breakpoint determination threshold on SURE on SIR}
	\label{fig:rq1_sir}
\end{figure}

\section{Result and Analysis}
\label{sect:result}

\subsection{RQ1: The value of the breakpoint determination threshold.}
\label{subsect:result_rq1}

As the first step, SURE needs to determine the Top-$x$\% riskiest program statements as breakpoints. In this RQ, we investigate the effectiveness of SURE when the breakpoint determination threshold is set from 0\% to 100\% with 10\% increments. Notice that in the training and deployment phases of SURE, feature extraction networks are necessary. Though RQ2 will investigate the selection of such networks, we must adopt a feature extraction network in this RQ to make the model run. To avoid bias, we will investigate this research question based on three feature extraction networks, i.e., VGG-16, AlexNet, and ResNet-18, separately. The results are given in Table~\ref{tab:result_rq1_sir} (in simulated environments) and Table~\ref{tab:result_rq1_d4j} (in real-world environments).

\subsubsection{In simulated environments}
\label{subsubsect:rq1_sir}

Let us first focus on the impact of the breakpoint determination threshold on the effectiveness of SURE, based on VGG-16 (i.e., the row ``VGG-16'' of Table~\ref{tab:result_rq1_sir}). In this table, each cell denotes  the value of a specific metric when using a variant of SURE. For example, the cell (``VGG-16'', ``10\%'', $V^T_{equal}$) is 161, indicating that when SURE takes the Top-10\% riskiest statements as breakpoints and adopts VGG-16 as the feature extraction network (SURE in such a configuration can be referred to as $SURE^{10\%}_{VGG-16}$, similarly below), it can correctly predict the number of faults on 161 faulty versions. Similarly, when $T$ takes $SURE^{10\%}_{VGG-16}$, the values of $S\_M^T_{FMI}$, $S\_M^T_{JC}$, $S\_M^T_{PR}$, and $S\_M^T_{RR}$, are 126.78, 106.39, 120.24, and 80.67, respectively. We can find that when we use VGG-16 as the feature extraction network, it does not matter much if the breakpoint determination threshold takes 10\%, 20\%, ..., or 100\%, because all the five metrics seem to be stable regardless of how many breakpoints are set. We illustrate this trend (i.e., the row ``VGG-16'' of Table~\ref{tab:result_rq1_sir}) in Figure~\ref{fig:rq1_sir}(a).

A similar trend can be found when AlexNet and ResNet-18 serve as the feature extraction network. Specifically, if we adopt AlexNet or ResNet-18 as the feature extraction network (i.e., the row ``AlexNet'' or the row ``ResNet-18'' of Table~\ref{tab:result_rq1_sir}), there is no explicit trend toward change in the five metrics with the increase of $x$ either, as shown in Figure~\ref{fig:rq1_sir}(b) and Figure~\ref{fig:rq1_sir}(c), respectively.

\begin{table}[t]
	\centering
	\renewcommand\arraystretch{1.17}
	\setlength{\belowcaptionskip}{3pt}
	\caption{\label{tab:result_rq1_d4j} Performance of SURE on Defects4J} 
	\resizebox{0.9\textwidth}{!}{
		\begin{tabular}{|cc|c|c|c|c|c|c|c|c|c|c|}
			\hline
			\rowcolor{gray!20} 
			\multicolumn{2}{|c|}{\cellcolor{gray!20}}                                   & \cellcolor{gray!20}\textbf{10\%}   & \cellcolor{gray!20}\textbf{20\%}   & \cellcolor{gray!20}\textbf{30\%}   & \cellcolor{gray!20}\textbf{40\%}   & \cellcolor{gray!20}\textbf{50\%}   & \cellcolor{gray!20}\textbf{60\%}   & \cellcolor{gray!20}\textbf{70\%}   & \cellcolor{gray!20}\textbf{80\%}   & \cellcolor{gray!20}\textbf{90\%}   & \cellcolor{gray!20}\textbf{100\%} \\ \hline
			
			\multicolumn{1}{|c|}{\cellcolor{gray!20}}                          & $\bm{V^T_{equal}}$ & \textbf{35}    & 37    & 31    & 30    & 31    & 37    & 36    & 28    & 32    & 31    \\ \cline{2-12} 
			\multicolumn{1}{|c|}{\cellcolor{gray!20}}                          & $\bm{S\_M^T_{FMI}}$    & \textbf{34.91} & 36.84 & 30.95 & 29.90  & 30.91 & 36.85 & 35.86 & 27.86 & 31.85 & 30.94 \\ \cline{2-12} 
			\multicolumn{1}{|c|}{\cellcolor{gray!20}}                          & $\bm{S\_M^T_{JC}}$     & \textbf{34.83} & 36.69 & 30.90  & 29.81 & 30.84 & 36.72 & 35.73 & 27.74 & 31.73 & 30.88 \\ \cline{2-12} 
			\multicolumn{1}{|c|}{\cellcolor{gray!20}}                          & $\bm{S\_M^T_{PR}}$     & \textbf{33.03} & 34.7  & 30.00    & 28.72 & 29.52 & 35.02 & 33.68 & 25.99 & 29.81 & 30.04 \\ \cline{2-12} 
			\multicolumn{1}{|c|}{\multirow{-5}{*}{\cellcolor{gray!20}\textbf{VGG-16}}}   & $\bm{S\_M^T_{RR}}$     & \textbf{33.05} & 34.41 & 29.92 & 28.58 & 29.37 & 34.67 & 33.51 & 25.85 & 29.48 & 29.95 \\ \hline

			\multicolumn{1}{|c|}{\cellcolor{gray!20}}                          & $\bm{V^T_{equal}}$ & \textbf{41}    & 40    & 32    & 34    & 31    & 38    & 34    & 32    & 33    & 32    \\ \cline{2-12} 
			\multicolumn{1}{|c|}{\cellcolor{gray!20}}                          & $\bm{S\_M^T_{FMI}}$   & \textbf{40.81} & 39.90  & 31.92 & 33.96 & 30.87 & 37.80  & 33.90  & 31.97 & 32.91 & 31.89 \\ \cline{2-12} 
			\multicolumn{1}{|c|}{\cellcolor{gray!20}}                          & $\bm{S\_M^T_{JC}}$     & \textbf{40.63} & 39.80  & 31.83 & 33.92 & 30.75 & 37.63 & 33.82 & 31.94 & 32.84 & 31.79 \\ \cline{2-12} 
			\multicolumn{1}{|c|}{\cellcolor{gray!20}}                          & $\bm{S\_M^T_{PR}}$     & \textbf{37.64} & 37.66 & 30.35 & 33.25 & 28.83 & 34.83 & 32.77 & 31.25 & 30.91 & 30.21 \\ \cline{2-12} 
			\multicolumn{1}{|c|}{\multirow{-5}{*}{\cellcolor{gray!20}\textbf{AlexNet}}} & $\bm{S\_M^T_{RR}}$     & \textbf{37.44} & 37.58 & 30.35 & 33.17 & 28.58 & 34.58 & 32.61 & 31.25 & 31.01 & 30.10  \\ \hline

			\multicolumn{1}{|c|}{\cellcolor{gray!20}}                          & $\bm{V^T_{equal}}$ & \textbf{37}    & 31    & 33    & 36    & 34    & 33    & 33    & 29    & 33    & 33    \\ \cline{2-12} 
			\multicolumn{1}{|c|}{\cellcolor{gray!20}}                          & $\bm{S\_M^T_{FMI}}$    & \textbf{36.85} & 30.90  & 32.91 & 35.92 & 33.88 & 32.89 & 32.92 & 28.88 & 32.90  & 32.93 \\ \cline{2-12} 
			\multicolumn{1}{|c|}{\cellcolor{gray!20}}                          & $\bm{S\_M^T_{JC}}$     & \textbf{36.71} & 30.82 & 32.83 & 35.84 & 33.78 & 32.81 & 32.85 & 28.79 & 32.80  & 32.88 \\ \cline{2-12} 
			\multicolumn{1}{|c|}{\cellcolor{gray!20}}                          & $\bm{S\_M^T_{PR}}$    & \textbf{34.36} & 29.41 & 31.07 & 34.04 & 32.03 & 30.83 & 31.75 & 27.04 & 31.16 & 32.04 \\ \cline{2-12} 
			\multicolumn{1}{|c|}{\multirow{-5}{*}{\cellcolor{gray!20}\textbf{ResNet-18}}}  & $\bm{S\_M^T_{RR}}$     & \textbf{34.20}  & 29.44 & 30.93 & 34.01 & 31.92 & 30.80  & 31.51 & 26.98 & 31.08 & 31.93 \\ \hline
	\end{tabular}}
\end{table}

\begin{figure}[t]
	\centering
	\includegraphics[width=0.9\linewidth]{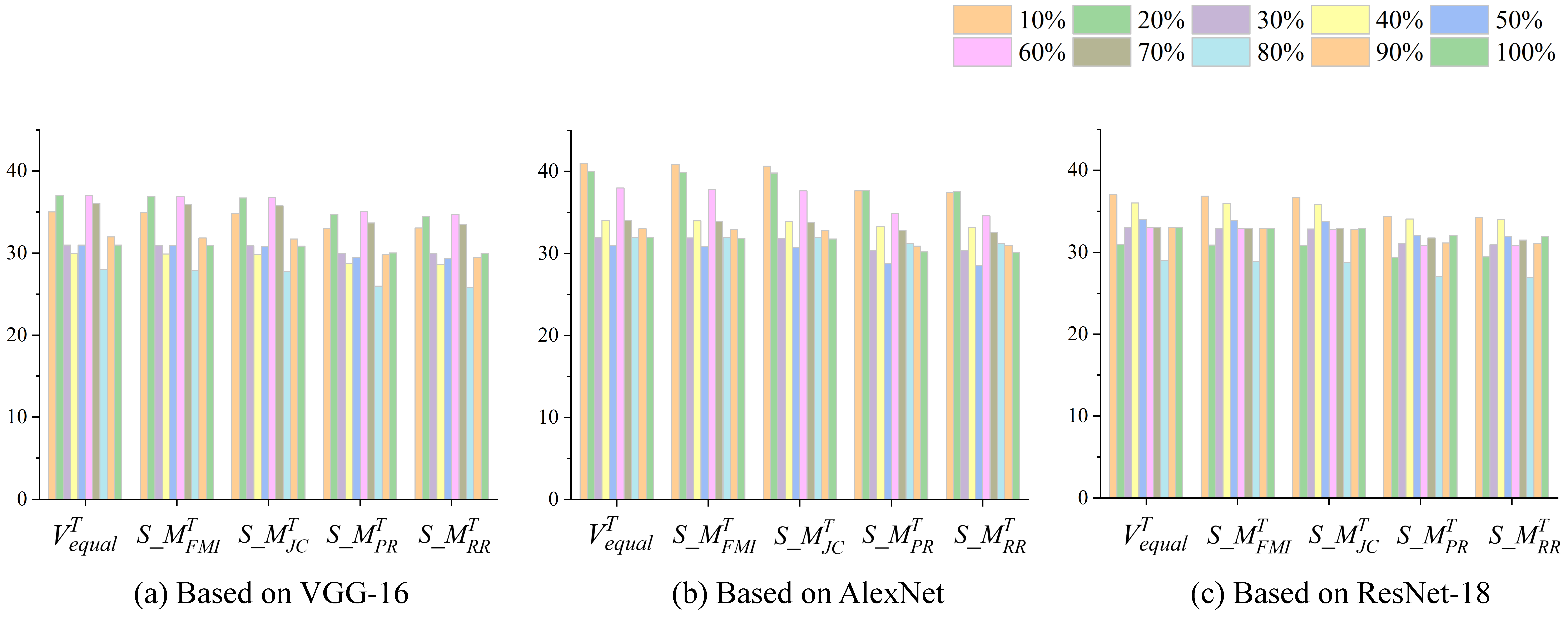}
	\caption{The impact of the breakpoint determination threshold on SURE on Defects4J}
	\label{fig:rq1_d4j}
\end{figure}

\subsubsection{In real-world environments}
\label{subsubsect:rq1_d4j}

Likewise, let us first use VGG-16 as the feature extraction network, to investigate the impact of the breakpoint determination threshold on the effectiveness of SURE. The results are reported in the row ``VGG-16'' of Table~\ref{tab:result_rq1_d4j}. Similar to the trend observed in simulated environments, we can find that the effectiveness of SURE is likely to be stable, regardless of how many breakpoints are set according to the suspiciousness of program statements. For example,  when the threshold is set from 10\% to 100\% with 10\% increments, the value of $V^T_{equal}$ is 35, 37, 31, 30, 31, 37, 36, 28, 32, and 31. And the values of $S\_M^T_{FMI}$, $S\_M^T_{JC}$,  $S\_M^T_{PR}$, and  $S\_M^T_{RR}$ seem to be not largely influenced by the breakpoint determination threshold either. We depict this trend in Figure~\ref{fig:rq1_d4j}(a). When the feature extraction model is replaced by AlexNet and ResNet-18, this conclusion can still be maintained, as shown in the row ``AlexNet'' and the row ``ResNet-18'' of Table~\ref{tab:result_rq1_d4j}, as well as Figure~\ref{fig:rq1_d4j}(b) and Figure~\ref{fig:rq1_d4j}(c), respectively.

The result of RQ1 can be explained to mean that the memory information collected at the Top-10\% riskiest statements is already sufficient to represent failures. This result confirms our intuition of the breakpoint determination phase (Section~\ref{subsect:breakpointdetermination}), i.e., statements with higher risk values are more likely to be faulty, and run-time memory information gathered at these positions could have a stronger capability to reveal faults, and thus can contribute more to representing failures. Moreover, the result of RQ1 also reveals the necessity of the breakpoint determination component of SURE. Specifically, more memory information cannot contribute more to failure representation, thus we can first use SBFL to determine a few breakpoints, and then collect memory information within a small range. This is practically important because the overhead of running SBFL is less than that of collecting run-time memory information. To put it another way, we preliminarily use a lower-cost technique to avoid the following procedure with relatively higher costs, while such a procedure could not bring gains for our approach. Therefore, \textbf{in the following RQs, the investigation will be based on the breakpoint determination threshold of Top-10\%}.

\subsection{RQ2: The adopted feature extraction network.}
\label{subsect:result_rq2}

In addition to the breakpoint determination threshold, another configurable hyperparameter of SURE is the feature extraction network. In this RQ, we select three deep convolutional neural networks (CNNs), VGG-16, AlexNet, and ResNet-18, to investigate how the effectiveness of SURE will be influenced when adopting different networks. The three investigated networks are all classical and commonly-used. Specifically, VGG-16 was introduced by the Visual Geometry Group (VGG) from the University of Oxford in 2014~\cite{simonyan2014very}, it is part of the VGG network series and is named ``16'' because it consists of 16 layers. VGG-16 is known for its simple yet effective architecture, which allows it to learn rich and complex features from images. Also, VGG-16's parameter count is relatively large thus it requires more computational resources. AlexNet is a pioneering deep CNN introduced by Krizhevsky et al. in 2012~\cite{krizhevsky2012imagenet}, which marks a significant breakthrough in computer vision. AlexNet popularized the use of ReLU activation functions, which helps alleviate the vanishing gradient problem and accelerate training. ResNet-18 is a variant of the residual neural network (ResNet) architecture. It was introduced as part of the ResNet family by He et al. in 2015~\cite{he2016deep}, which uses residual blocks to address the vanishing gradient problem that often occurs in deep neural networks. In this RQ, we analyze how the effectiveness of SURE will be influenced when adopting  VGG-16, AlexNet, and ResNet, based on the run-time memory information collected at the Top-10\% riskiest statement.

\subsubsection{In simulated environments}
\label{subsubsect:rq2_sir}

\begin{figure}[t]
	\centering
	\includegraphics[width=0.85\linewidth]{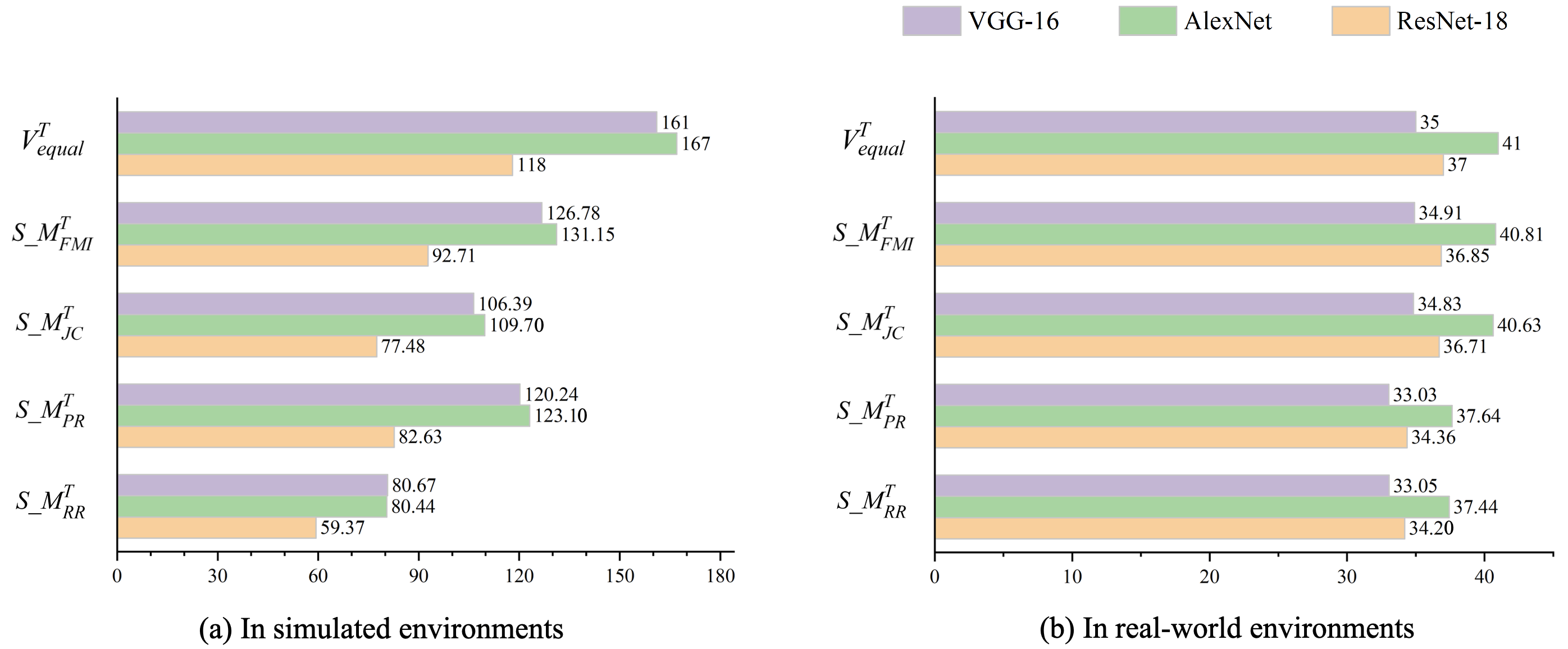}
	\caption{The impact of the feature extraction network on SURE}
	\label{fig:rq2}
\end{figure}

We first analyze the results in simulated environments, as shown in the column ``10\%'' of Table~\ref{tab:result_rq1_sir} (This column is in bold). We can find that there are 167 ``$k$ == $r$'' faulty versions when using AlexNet as the feature extraction network, while 161 and 118 ones when using VGG-16 and ResNet-18, respectively. As for the metrics $S\_M^T_{FMI}$, $S\_M^T_{JC}$, $S\_M^T_{PR}$, and $S\_M^T_{RR}$, AlexNet also dominates VGG-16 and ResNet-18 (with only one exception where $S\_M^{SURE^{10\%}_{VGG-16}}_{RR}$: 80.67 is slightly higher than $S\_M^{SURE^{10\%}_{AlexNet}}_{RR}$: 80.44). We illustrate these results in Figure~\ref{fig:rq2}(a) as well for clear understanding. We can find that in simulated environments, AlexNet can better extract features for PMS images thus deliver more promising failure indexing outcomes.

\subsubsection{In real-world environments}
\label{subsubsect:rq2_d4j}

Now we navigate to real-world environments, the results are given in the column ``10\%'' of Table~\ref{tab:result_rq1_d4j} (This column is in bold). We can find that there are 41 ``$k$ == $r$'' faulty versions when using AlexNet as the feature extraction network, while 35 and 37 ones when using VGG-16 and ResNet-18, respectively. As for the metrics $S\_M^T_{FMI}$, $S\_M^T_{JC}$, $S\_M^T_{PR}$, and $S\_M^T_{RR}$, AlexNet also dominates VGG-16 and ResNet-18. We illustrate these results in Figure~\ref{fig:rq2}(b) as well for clear understanding. Similar to the conclusion in simulated environments, we can find that AlexNet is more suitable for extracting features for PMS images.

 A possible explanation for the promise of AlexNet in our experiments could be its relatively shallow architecture. Specifically, PMS images are typically in a simple form (not too many complex features involved in PMS images). It can be naturally conjectured that a shallow CNN network might be enough to extract their features. Among the selected three networks, AlexNet consists of 8 layers, including 5 convolutional layers and 3 fully connected layers, while VGG-16 and ResNet-18 both have much deeper architectures than AlexNet. Thus, it is intuitive that SURE is more suitable to configure AlexNet. Therefore, \textbf{in the following RQs, the investigation will be based on the feature extraction network of AlexNet}. 
 
\subsection{RQ3: The competitiveness of SURE.}
\label{subsect:result_rq3}

In this RQ, we investigate to which extent SURE exceeds the existing most prevalent and advanced failure indexing techniques. Given the conclusions of RQ1 and RQ2, we configure 10\% as the breakpoint determination threshold and AlexNet as the feature extraction network for SURE in this RQ. To put it another way, we will use $SURE^{10\%}_{AlexNet}$ to compare with baseline techniques (for convenience, we use the term ``$SURE$'' to denote ``$SURE^{10\%}_{AlexNet}$'' hereafter).

\subsubsection{In simulated environments}
\label{subsubsect:rq3_sir}

% 这张表里附有SURE相较于每个baseline的提升幅度
\iffalse
\begin{table}\small
	\caption{Competitiveness of SURE in simulated environments}
	\label{tab:result_rq3_sir}
	\begin{tabular}{p{1.5cm}p{1.7cm}<{\centering}p{2.1cm}<{\centering}p{2.1cm}<{\centering}p{2.1cm}<{\centering}p{2.1cm}<{\centering}}
		\toprule
		\centering \textbf{T} & \bm{$V^T_{equal}$} & \bm{$S\_M^T_{FMI}$}& \bm{$S\_M^T_{JC}$} & \bm{$S\_M^T_{PR}$} & \bm{$S\_M^T_{RR}$} \\
		\midrule
		$Cov_{hit}$ & 32 (421.88\%)& 25.31 (418.17\%)& 21.34 (414.06\%)& 23.18 (431.06\%)& 14.92 (439.14\%) \\
		$Cov_{count}$ & 54 (209.26\%) & 43.33 (202.68\%)& 36.75 (198.50\%)& 42.37 (190.54\%)& 31.30 (157.00\%)\\
		$MSeer$ & \textbf{83 (101.20\%)}& \textbf{62.86 (108.64\%)} & \textbf{51.72 (112.10\%)}& \textbf{63.08 (95.15\%)}& \textbf{39.26 (104.89\%)}\\
		$MSeer_{GP19}$ & 128 (30.47\%) & 99.98 (31.18\%) & 84.64 (29.61\%)& 104.09 (18.26\%)& 73.38 (9.62\%) \\
		$SURE$ & \textbf{167}& \textbf{131.15} & \textbf{109.70} & \textbf{123.10} & \textbf{80.44} \\

		\bottomrule
	\end{tabular}
\end{table}
\fi

The results in simulated environments are given in Table~\ref{tab:result_rq3_sir} and Figure~\ref{fig:rq3}(a).  It can be observed that $SURE$ outperforms all the baseline techniques substantially. Specifically, regarding the capability of faults number estimation, $SURE$ can correctly predict the number of faults on $167$ faulty versions on SIR, with 421.88\%, 209.26\%, 101.20\%, and 30.47\% improvements compared with $Cov_{hit}$ (32), $Cov_{count}$ (54), $MSeer$ (83), and $MSeer_{GP19}$ (128),  respectively. $SURE$ consistently exceeds the baselines on all four clustering metrics. For instance, if we focus on the comparison between $SURE$ and $Cov_{hit}$, improvements are 418.17\%, 414.06\%, 431.06\%, and 439.14\%, regarding $S\_M^T_{FMI}$, $S\_M^T_{JC}$, $S\_M^T_{PR}$, and $S\_M^T_{RR}$ , respectively. Considering the four clustering metrics globally, the average improvement of $SURE$ over $Cov_{hit}$ is 425.61\%. Similarly, in the contexts of comparing $SURE$ with $Cov_{count}$, $MSeer$,  and $MSeer_{GP19}$, the average improvements can be calculated as 187.18\% and 105.20\%, and 22.17\% respectively.

\iffalse
Even if comparing with $MSeer_{GP19}$, an enhanced variant of MSeer, $SURE$ is also highly-competitive: $SURE$ can correctly predict the number of faults on 30.47\% more faulty versions than $MSeer$, and the global improvement of the four clustering metrics is 22.17\%.
\fi

\subsubsection{In real-world environments}
\label{subsubsect:rq3_d4j}

\iffalse
\begin{table}\small
	\caption{Competitiveness of SURE in real-world environments}
	\label{tab:result_rq3_d4j}
	\begin{tabular}{p{1.5cm}p{1.7cm}<{\centering}p{2.1cm}<{\centering}p{2.1cm}<{\centering}p{2.1cm}<{\centering}p{2.1cm}<{\centering}}
		\toprule
		\centering \textbf{T} & \bm{$V^T_{equal}$} & \bm{$S\_M^T_{FMI}$} & \bm{$S\_M^T_{JC}$} & \bm{$S\_M^T_{PR}$} & \bm{$S\_M^T_{RR}$} \\
		\midrule
		$Cov_{hit}$ & 20 (105.00\%)& 20.00 (104.05\%)& 20.00 (103.15\%)& 20.00 (88.20\%)& 20.00 (87.20\%) \\
		$Cov_{count}$ & 24 (70.83\%) & 23.98 (70.18\%)& 23.96 (69.57\%)& 23.50 (60.17\%)& 23.50 (59.32\%)\\
		$MSeer$ & \textbf{29 (41.38\%)}& \textbf{28.99 (40.77\%)} & \textbf{28.98 (40.20\%)}& \textbf{28.75 (30.92\%)}& \textbf{28.75 (30.23\%)}\\
		$MSeer_{GP19}$ & 32 (28.13\%) & 31.98 (27.61\%) & 31.95 (27.17\%)& 31.25 (20.45\%)& 31.25 (19.81\%) \\
		$SURE$ & \textbf{41}& \textbf{40.81} & \textbf{40.63} & \textbf{37.64} & \textbf{37.44} \\
		
		\bottomrule
	\end{tabular}
\end{table}
\fi

\begin{table}\small
	\caption{Competitiveness of SURE in simulated environments}
	\label{tab:result_rq3_sir}
	\begin{tabular}{p{1.5cm}p{1.3cm}<{\centering}p{1.6cm}<{\centering}p{1.6cm}<{\centering}p{1.6cm}<{\centering}p{1.6cm}<{\centering}}
		\toprule
		\centering \textbf{T} & \bm{$V^T_{equal}$} & \bm{$S\_M^T_{FMI}$}& \bm{$S\_M^T_{JC}$} & \bm{$S\_M^T_{PR}$} & \bm{$S\_M^T_{RR}$} \\
		\midrule
		$Cov_{hit}$ & 32 & 25.31 & 21.34 & 23.18& 14.92 \\
		$Cov_{count}$ & 54 & 43.33 & 36.75 & 42.37& 31.30 \\
		$MSeer$ & 83& 62.86  & 51.72 & 63.08 & 39.26\\
		$MSeer_{GP19}$ & 128  & 99.98 & 84.64& 104.09 & 73.38  \\
		\bm{$SURE$} & \textbf{167}& \textbf{131.15} & \textbf{109.70} & \textbf{123.10} & \textbf{80.44} \\
		
		\bottomrule
	\end{tabular}
\end{table}

\begin{table}\small
	\caption{Competitiveness of SURE in real-world environments}
	\label{tab:result_rq3_d4j}
	\begin{tabular}{p{1.5cm}p{1.3cm}<{\centering}p{1.6cm}<{\centering}p{1.6cm}<{\centering}p{1.6cm}<{\centering}p{1.6cm}<{\centering}}
		\toprule
		\centering \textbf{T} & \bm{$V^T_{equal}$} & \bm{$S\_M^T_{FMI}$} & \bm{$S\_M^T_{JC}$} & \bm{$S\_M^T_{PR}$} & \bm{$S\_M^T_{RR}$} \\
		\midrule
		$Cov_{hit}$ & 20 & 20.00 & 20.00 & 20.00 & 20.00  \\
		$Cov_{count}$ & 24  & 23.98 & 23.96 & 23.50 & 23.50 \\
		$MSeer$ & 29& 28.99  & 28.98 & 28.75& 28.75\\
		$MSeer_{GP19}$ & 32  & 31.98  & 31.95 & 31.25& 31.25 \\
		\bm{$SURE$} & \textbf{41}& \textbf{40.81} & \textbf{40.63} & \textbf{37.64} & \textbf{37.44} \\
		
		\bottomrule
	\end{tabular}
\end{table}

\begin{figure}[t]
	\centering
	\includegraphics[width=0.85\linewidth]{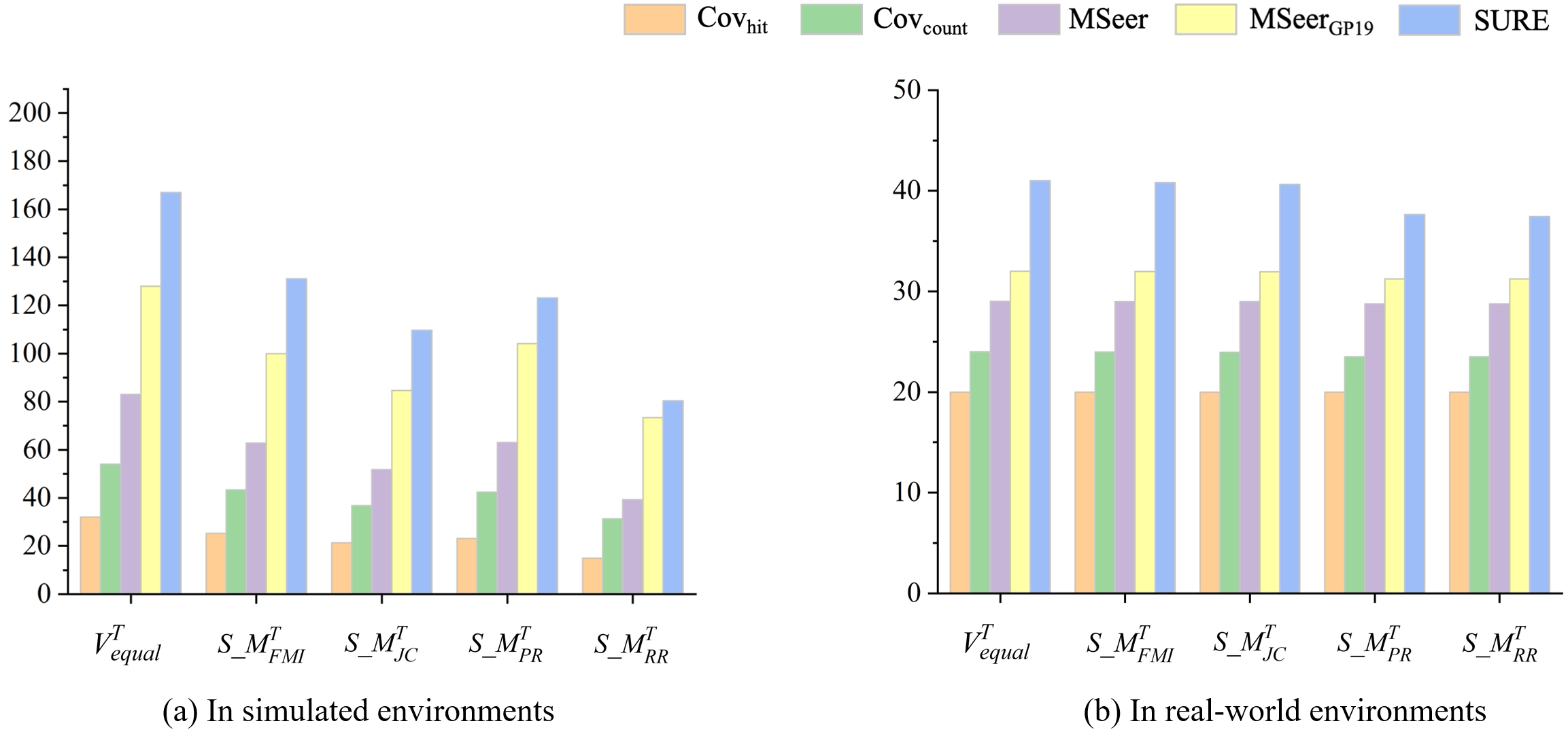}
	\caption{Competitiveness of SURE}
	\label{fig:rq3}
\end{figure}

The results in simulated environments are given in Table~\ref{tab:result_rq3_d4j} and Figure~\ref{fig:rq3}(b).  It can be observed that $SURE$ outperforms all the baseline techniques substantially. Specifically, regarding the capability of faults number estimation, $SURE$ can correctly predict the number of faults on $41$ faulty versions on Defects4J, with 105.00\%, 70.83\%, 41.38\%, and 28.13\% improvements compared with $Cov_{hit}$ (20), $Cov_{count}$ (24), $MSeer$ (29), and $MSeer_{GP19}$ (32), respectively. $SURE$ consistently exceeds the baselines on all four clustering metrics. For instance, if we focus on the comparison between $SURE$ and $Cov_{hit}$, improvements are 104.05\%, 103.15\%, 88.20\%, and 87.20\%, regarding $S\_M^T_{FMI}$, $S\_M^T_{JC}$, $S\_M^T_{PR}$, and $S\_M^T_{RR}$ , respectively. Considering the four clustering metrics globally, the average improvement of $SURE$ over $Cov_{hit}$ is 95.65\%. Similarly, in the contexts of comparing $SURE$ with $Cov_{Count}$, $MSeer$, and $MSeer_{GP19}$,  the average improvements can be calculated as 64.81\%, 35.53\%, and 23.76\%, respectively.

\iffalse
Even if comparing with $MSeer_{GP19}$, an enhanced variant of MSeer, $SURE$ is also highly-competitive: $SURE$ can correctly predict the number of faults on 28.13\% more faulty versions than $MSeer$, and the global improvement of the four clustering metrics is 23.76\%.
\fi

\syrevise{As a reminder, MSeer is the state-of-the-art technique in the field of failure indexing. MSeer$_{\rm{GP19}}$ is not an existing published failure indexing technique, it is manually created by us to further evaluate the competitiveness of SURE.}

\subsection{RQ4: The comprehensibility of SURE.}
\label{subsect:result_rq4}

In Section~\ref{sect:introduction}, we point out that one of the two longstanding challenges in failure indexing is the lack of \syrevise{comprehensibility}, that is, the result of failure indexing is hard to \syrevise{comprehend} for human developers, which could prevent the result from being applied. The \syrevise{comprehensibility} of a failure indexing technique essentially lies in the \syrevise{comprehensibility} of the failure proximity, because it is the failure proximity that represents failures and hence provides human developers with the evidence of failure indexing. Here we describe the \syrevise{comprehensibility} of failure indexing techniques as follows:

\vspace{0.2cm}
\begin{center}
	\begin{tcolorbox}[colback=gray!10,%gray background
		colframe=black,% black frame colour
		width=13cm,% Use 8cm total width,
		height=1.85cm,
		arc=1mm, auto outer arc,
		boxrule=0.5pt,
		]
		\vspace{-0.15cm}\emph{A failure indexing technique is considered to have strong \syrevise{comprehensibility}, if given the failure indexing result as well as the evidence (i.e., the proxies for failed test cases) for how this result was obtained, \textbf{developers can be immediately convinced by the result based on the evidence}.}
	\end{tcolorbox}
\end{center}

\begin{figure}[t]
	\centering
	\includegraphics[width=0.85\linewidth]{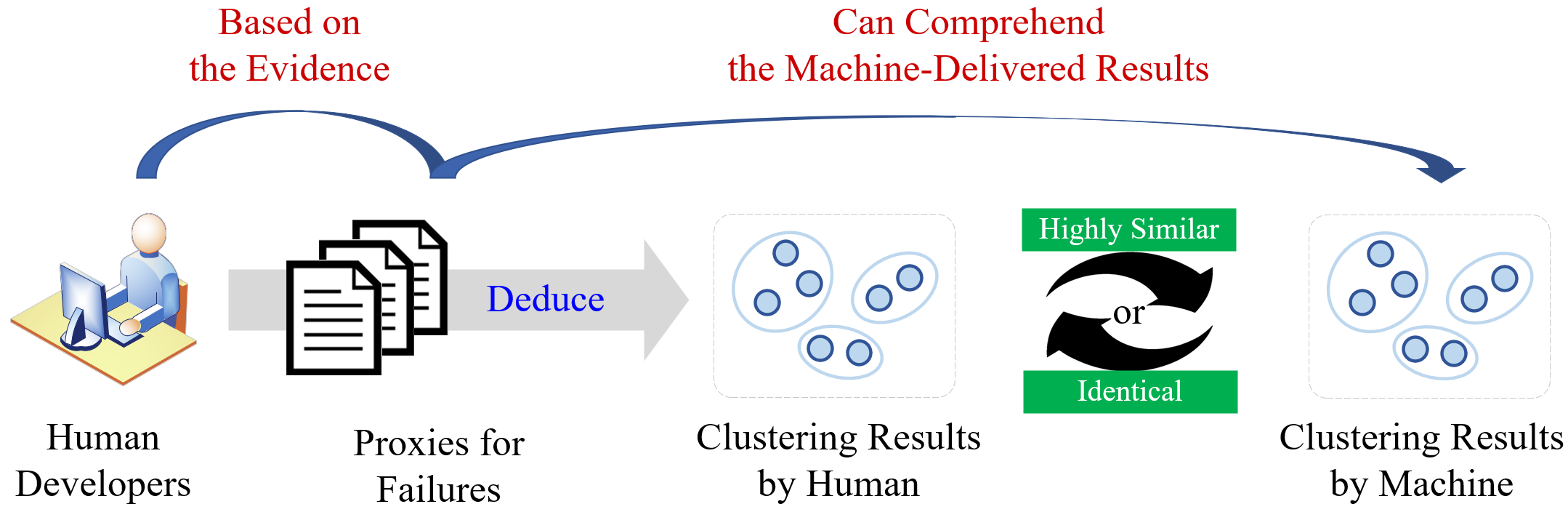}
	\caption{Quantitatively measure the \syrevise{comprehensibility} of failure indexing techniques}
	\label{fig:human_study}
\end{figure}

But this description cannot directly guide the investigation of RQ4, because it is subjective for developers to determine whether they are convinced by the result or not, and it is difficult to quantitatively measure the extent to which they are convinced. To tackle this problem, we propose the following strategy to quantify the \syrevise{comprehensibility} of a failure indexing technique:

\vspace{0.2cm}
\begin{center}
	\begin{tcolorbox}[colback=gray!10,%gray background
		colframe=black,% black frame colour
		width=13cm,% Use 8cm total width,
		height=2.3cm,
		arc=1mm, auto outer arc,
		boxrule=0.5pt,
		]
		\vspace{-0.15cm}\emph{Providing developers with only the evidence (i.e., the proxies for failed test cases) of a failure indexing process, and letting them manually cluster failed test cases according to the evidence. \textbf{The closer their manual clustering outcomes are to the failure indexing results, the better they \syrevise{comprehend} the evidence}, and thus, this failure indexing technique has stronger \syrevise{comprehensibility}.}
	\end{tcolorbox}
\end{center}
The intuition behind this strategy can be described in Figure~\ref{fig:human_study}. Specifically, if only  the evidence of a failure indexing process is provided to developers, and the result of their manual clustering based on that evidence is similar or even identical to the result of the failure indexing technique, then when the failure indexing result as well as the evidence are provided together, they are able to \syrevise{comprehend} the evidence.

As such, the clustering effectiveness made by human developers can serve as the \syrevise{comprehensibility} of a failure indexing technique. Hence, the evaluation metrics introduced in Section~\ref{subsect:metric}, i.e., $V^T_{equal}$, $S\_M^T_{FMI}$, $S\_M^T_{JC}$, $S\_M^T_{PR}$, and $S\_M^T_{RR}$, can be directly employed in this RQ.

We recruit 15 participants from \syrevise{Wuhan University}, all of whom are graduate students in Computer Science and have at least four years of programming experience. They are requested to manually finish 9 failure indexing tasks (each task corresponds to an $r$-bug faulty program, where $r$ = 1, 2, 3, 4, or 5). Specifically, in each task, only a series of failed test cases are provided to participants. These failures are triggered by how many underlying fault(s), as well as the mutual relationship among these failures, are hidden from participants. Participants will manually divide all failures through two steps: 1) Estimating these failures should be divided into how many clusters, i.e., there are how many underlying faults behind these failures. 2) Manually dividing failures according to the estimated number of clusters.

As for each task, its failed test cases are given in one of three forms of representation: program memory spectrum (PMS) images, suspiciousness ranking lists by GP19, and coverage vectors of execution frequency. Specifically, PMS images are the new failure representation proposed in this paper, suspiciousness ranking lists by GP19 are the failure representation of $MSeer_{GP19}$, and coverage vectors of execution frequency are the failure representation of $Cov_{count}$. We do not investigate suspiciousness ranking lists by Crosstab (the failure representation of $MSeer$), and coverage vectors of binary indicators (the failure representation of $Cov_{hit}$). This is because $MSeer$ and $MSeer_{GP19}$ both belong to the SD-based failure proximity, while $MSeer_{GP19}$ has been demonstrated to be better than $MSeer$ in Section~\ref{subsect:result_rq3}. Also, $Cov_{hit}$ and $Cov_{count}$ both belong to the CC-based failure proximity, while $Cov_{count}$ has been demonstrated to be better than $Cov_{hit}$ in Section~\ref{subsect:result_rq3}.

% \usepackage{multirow}
\begin{table}[]
	\caption{Nine failure indexing tasks}
	\label{tab:9tasks}
	\begin{threeparttable}
		\begin{tabular}{ccccccc}
			\hline
			\multirow{2}{*}{\textbf{Project}} & \multirow{2}{*}{\textbf{Language}} & \multirow{2}{*}{\textbf{\# Failures}} & \multirow{2}{*}{\textbf{\# Faults}} & \multicolumn{3}{c}{\textbf{Failre representation for developers}}              \\ \cline{5-7} 
			&                                    &                                       &                                     & PMS                      & Ranking list*             & Coverage\footnotesize$^\dagger$                 \\ \hline
			Task-1                            & C                                  & 10                                    & 1                                   & \textbf{1}, 4, 7, 10, 13 & 2, 5, 8, 11, 14          & 3, 6, 9, 12, 15          \\
			Task-2                            & C                                  & 29                                    & 2                                   & \textbf{1}, 4, 7, 10, 13 & 2, 5, 8, 11, 14          & 3, 6, 9, 12, 15          \\
			Task-3                            & C                                  & 10                                    & 3                                   & \textbf{1}, 4, 7, 10, 13 & 2, 5, 8, 11, 14          & 3, 6, 9, 12, 15          \\
			Task-4                            & C                                  & 19                                    & 4                                   & 3, 6, 9, 12, 15          & \textbf{1}, 4, 7, 10, 13& 2, 5, 8, 11, 14          \\
			Task-5                            & C                                  & 17                                    & 5                                   & 3, 6, 9, 12, 15          & \textbf{1}, 4, 7, 10, 13& 2, 5, 8, 11, 14          \\
			Task-6                            & Java                               & 6                                     & 2                                   & 3, 6, 9, 12, 15          & \textbf{1}, 4, 7, 10, 13 & 2, 5, 8, 11, 14          \\
			Task-7                            & Java                               & 3                                     & 3                                   & 2, 5, 8, 11, 14          & 3, 6, 9, 12, 15          & \textbf{1}, 4, 7, 10, 13 \\
			Task-8                            & Java                               & 4                                     & 4                                   & 2, 5, 8, 11, 14          & 3, 6, 9, 12, 15          & \textbf{1}, 4, 7, 10, 13\\
			Task-9                            & Java                               & 5                                     & 5                                   & 2, 5, 8, 11, 14          & 3, 6, 9, 12, 15          & \textbf{1}, 4, 7, 10, 13\\ \hline
		\end{tabular}
		
		\begin{tablenotes}
			\footnotesize
			\item * ``Ranking list'' denotes ``Suspiciousness ranking lists by GP19''. The same in Table~\ref{tab:result_rq4}.
			\item $\dagger$ ``Coverage'' denotes ``Coverage vectors of execution frequency''. The same in Table~\ref{tab:result_rq4}.
		\end{tablenotes}
		
	\end{threeparttable}
\end{table}

We concisely introduce the 9 tasks in Table~\ref{tab:9tasks}, including the programming language, the number of observed failures in testing, as well as the number of underlying faults. \textbf{From the perspective of tasks}, each task will be carried out by all of the 15 participants, among them, five are based on PMS, five are based on suspiciousness ranking lists by GP19, and five are based on coverage vectors of execution frequency. For example, we can find that for Task-1, Task-2, and Task-3, Participants 1, 4, 7, 10, 13 will be based on PMS to perform failure indexing, Participants 2, 5, 8, 11, 14 will be based on suspiciousness ranking lists by GP19, and Participants 3, 6, 9, 12, 15 will be based on coverage vectors of execution frequency. Additionally, \textbf{from the perspective of participants}, each developer will handle all of the 9 tasks, among them, three for each of the three forms of  representation. For example, we can find that Participant-1 handles Task-1, Task-2, and Task-3 based on PMS, handles Task-4, Task-5, and Task-6 based on suspiciousness ranking lists by GP19, and handles Task-7, Task-8, and Task-9 based on coverage vectors of execution frequency, as shown by the bold numbers in Table~\ref{tab:9tasks}. As such, each form of representation will be carried out for 45 times, as shown in the columns ``PMS'', ``Ranking list'', and ``Coverage'' in Table~\ref{tab:9tasks}.

\begin{table}[]
	\centering
	\renewcommand\arraystretch{1.2}
	\setlength{\belowcaptionskip}{3pt}
	\caption{\label{tab:result_rq4} Comprehensibility of three forms of failure representation} 
	\resizebox{0.95\textwidth}{!}{
	\begin{tabular}{|cc|ccccccccccccccc|c|}
		\hline
		\multicolumn{2}{|c|}{}                                                                    & \multicolumn{15}{c|}{\textbf{Participant}}                                                                                                                                                                                                                                                                                                                                                                                                                                                                                                                                                                                                                                                                                                                                                         &                                  \\ \cline{3-17}
		\multicolumn{2}{|c|}{\multirow{-2}{*}{\textbf{}}}                                         & \multicolumn{1}{c|}{No. 1}                         & \multicolumn{1}{c|}{No. 2}                         & \multicolumn{1}{c|}{No. 3}                         & \multicolumn{1}{c|}{No. 4}                         & \multicolumn{1}{c|}{\textbf{No. 5}}                         & \multicolumn{1}{c|}{No. 6}                         & \multicolumn{1}{c|}{No. 7}                         & \multicolumn{1}{c|}{No. 8}                         & \multicolumn{1}{c|}{No. 9}                         & \multicolumn{1}{c|}{No. 10}                        & \multicolumn{1}{c|}{No. 11}                        & \multicolumn{1}{c|}{No. 12}                        & \multicolumn{1}{c|}{No. 13}                       & \multicolumn{1}{c|}{No. 14}                        & No. 15                        & \multirow{-2}{*}{\textbf{Total}} \\ \hline
		\multicolumn{1}{|c|}{}                                        & \bm{$V^T_{equal}$}    & \multicolumn{1}{c|}{2}                             & \multicolumn{1}{c|}{3}                             & \multicolumn{1}{c|}{2}                             & \multicolumn{1}{c|}{3}                             & \multicolumn{1}{c|}{\textbf{3}}                             & \multicolumn{1}{c|}{3}                             & \multicolumn{1}{c|}{3}                             & \multicolumn{1}{c|}{3}                             & \multicolumn{1}{c|}{3}                             & \multicolumn{1}{c|}{2}                             & \multicolumn{1}{c|}{3}                             & \multicolumn{1}{c|}{2}                             & \multicolumn{1}{c|}{1}                            & \multicolumn{1}{c|}{3}                             & 3                             & 39                               \\ \cline{2-18} 
		\multicolumn{1}{|c|}{}                                        & \bm{$S\_M^T_{FMI}$}   & \multicolumn{1}{c|}{2.00}                          & \multicolumn{1}{c|}{3.00}                          & \multicolumn{1}{c|}{2.00}                          & \multicolumn{1}{c|}{3.00}                          & \multicolumn{1}{c|}{\textbf{3.00}}                          & \multicolumn{1}{c|}{3.00}                          & \multicolumn{1}{c|}{3.00}                          & \multicolumn{1}{c|}{3.00}                          & \multicolumn{1}{c|}{3.00}                          & \multicolumn{1}{c|}{2.00}                          & \multicolumn{1}{c|}{3.00}                          & \multicolumn{1}{c|}{2.00}                          & \multicolumn{1}{c|}{1.00}                         & \multicolumn{1}{c|}{3.00}                          & 3.00                          & 39.00                            \\ \cline{2-18} 
		\multicolumn{1}{|c|}{}                                        & \bm{$S\_M^T_{JC}$}    & \multicolumn{1}{c|}{2.00}                          & \multicolumn{1}{c|}{3.00}                          & \multicolumn{1}{c|}{2.00}                          & \multicolumn{1}{c|}{3.00}                          & \multicolumn{1}{c|}{\textbf{3.00}}                          & \multicolumn{1}{c|}{3.00}                          & \multicolumn{1}{c|}{3.00}                          & \multicolumn{1}{c|}{3.00}                          & \multicolumn{1}{c|}{3.00}                          & \multicolumn{1}{c|}{2.00}                          & \multicolumn{1}{c|}{3.00}                          & \multicolumn{1}{c|}{2.00}                          & \multicolumn{1}{c|}{1.00}                         & \multicolumn{1}{c|}{3.00}                          & 3.00                          & 39.00                            \\ \cline{2-18} 
		\multicolumn{1}{|c|}{}                                        & \bm{$S\_M^T_{PR}$}    & \multicolumn{1}{c|}{2.00}                          & \multicolumn{1}{c|}{3.00}                          & \multicolumn{1}{c|}{2.00}                          & \multicolumn{1}{c|}{3.00}                          & \multicolumn{1}{c|}{\textbf{3.00}}                          & \multicolumn{1}{c|}{3.00}                          & \multicolumn{1}{c|}{3.00}                          & \multicolumn{1}{c|}{3.00}                          & \multicolumn{1}{c|}{3.00}                          & \multicolumn{1}{c|}{2.00}                          & \multicolumn{1}{c|}{3.00}                          & \multicolumn{1}{c|}{1.96}                          & \multicolumn{1}{c|}{1.00}                         & \multicolumn{1}{c|}{3.00}                          & 3.00                          & 38.96                            \\ \cline{2-18} 
		\multicolumn{1}{|c|}{}                                        & \bm{$S\_M^T_{RR}$}    & \multicolumn{1}{c|}{2.00}                          & \multicolumn{1}{c|}{3.00}                          & \multicolumn{1}{c|}{2.00}                          & \multicolumn{1}{c|}{3.00}                          & \multicolumn{1}{c|}{\textbf{3.00}}                          & \multicolumn{1}{c|}{3.00}                          & \multicolumn{1}{c|}{3.00}                          & \multicolumn{1}{c|}{3.00}                          & \multicolumn{1}{c|}{3.00}                          & \multicolumn{1}{c|}{2.00}                          & \multicolumn{1}{c|}{3.00}                          & \multicolumn{1}{c|}{1.93}                          & \multicolumn{1}{c|}{1.00}                         & \multicolumn{1}{c|}{3.00}                          & 3.00                          & 38.93                            \\ \cline{2-18} 
		\multicolumn{1}{|c|}{\multirow{-6}{*}{\textbf{PMS}}}          & \textbf{Average Time (s)} & \multicolumn{1}{c|}{\cellcolor{gray!20}5.41}  & \multicolumn{1}{c|}{\cellcolor{gray!20}3.08}  & \multicolumn{1}{c|}{\cellcolor{gray!20}1.50}  & \multicolumn{1}{c|}{\cellcolor{gray!20}6.22}  & \multicolumn{1}{c|}{\cellcolor{gray!20}\textbf{1.17}}  & \multicolumn{1}{c|}{\cellcolor{gray!20}3.24}  & \multicolumn{1}{c|}{\cellcolor{gray!20}1.39}  & \multicolumn{1}{c|}{\cellcolor{gray!20}3.42}  & \multicolumn{1}{c|}{\cellcolor{gray!20}4.43}  & \multicolumn{1}{c|}{\cellcolor{gray!20}5.63}  & \multicolumn{1}{c|}{\cellcolor{gray!20}3.11}  & \multicolumn{1}{c|}{\cellcolor{gray!20}2.62}  & \multicolumn{1}{c|}{\cellcolor{gray!20}3.02} & \multicolumn{1}{c|}{\cellcolor{gray!20}2.33}  & \cellcolor{gray!20}2.33  & \cellcolor{gray!20}3.26     \\ \hline
		\multicolumn{1}{|c|}{}                                        & \bm{$V^T_{equal}$}    & \multicolumn{1}{c|}{2}                             & \multicolumn{1}{c|}{0}                             & \multicolumn{1}{c|}{3}                             & \multicolumn{1}{c|}{2}                             & \multicolumn{1}{c|}{\textbf{0}}                             & \multicolumn{1}{c|}{1}                             & \multicolumn{1}{c|}{0}                             & \multicolumn{1}{c|}{1}                             & \multicolumn{1}{c|}{2}                             & \multicolumn{1}{c|}{2}                             & \multicolumn{1}{c|}{1}                             & \multicolumn{1}{c|}{1}                             & \multicolumn{1}{c|}{1}                            & \multicolumn{1}{c|}{0}                             & 3                             & 19                               \\ \cline{2-18} 
		\multicolumn{1}{|c|}{}                                        & \bm{$S\_M^T_{FMI}$}   & \multicolumn{1}{c|}{1.99}                          & \multicolumn{1}{c|}{0.00}                          & \multicolumn{1}{c|}{3.00}                          & \multicolumn{1}{c|}{1.99}                          & \multicolumn{1}{c|}{\textbf{0.00}}                          & \multicolumn{1}{c|}{1.00}                          & \multicolumn{1}{c|}{0.00}                          & \multicolumn{1}{c|}{0.99}                          & \multicolumn{1}{c|}{2.00}                          & \multicolumn{1}{c|}{1.99}                          & \multicolumn{1}{c|}{0.99}                          & \multicolumn{1}{c|}{1.00}                          & \multicolumn{1}{c|}{1.00}                         & \multicolumn{1}{c|}{0.00}                          & 3.00                          & 18.95                            \\ \cline{2-18} 
		\multicolumn{1}{|c|}{}                                        & \bm{$S\_M^T_{JC}$}    & \multicolumn{1}{c|}{1.99}                          & \multicolumn{1}{c|}{0.00}                          & \multicolumn{1}{c|}{3.00}                          & \multicolumn{1}{c|}{1.99}                          & \multicolumn{1}{c|}{\textbf{0.00}}                          & \multicolumn{1}{c|}{1.00}                          & \multicolumn{1}{c|}{0.00}                          & \multicolumn{1}{c|}{0.98}                          & \multicolumn{1}{c|}{2.00}                          & \multicolumn{1}{c|}{1.99}                          & \multicolumn{1}{c|}{0.98}                          & \multicolumn{1}{c|}{1.00}                          & \multicolumn{1}{c|}{1.00}                         & \multicolumn{1}{c|}{0.00}                          & 3.00                          & 18.93                            \\ \cline{2-18} 
		\multicolumn{1}{|c|}{}                                        & \bm{$S\_M^T_{PR}$}    & \multicolumn{1}{c|}{1.81}                          & \multicolumn{1}{c|}{0.00}                          & \multicolumn{1}{c|}{3.00}                          & \multicolumn{1}{c|}{1.62}                          & \multicolumn{1}{c|}{\textbf{0.00}}                          & \multicolumn{1}{c|}{1.00}                          & \multicolumn{1}{c|}{0.00}                          & \multicolumn{1}{c|}{0.46}                          & \multicolumn{1}{c|}{2.00}                          & \multicolumn{1}{c|}{1.73}                          & \multicolumn{1}{c|}{0.55}                          & \multicolumn{1}{c|}{1.00}                          & \multicolumn{1}{c|}{1.00}                         & \multicolumn{1}{c|}{0.00}                          & 3.00                          & 17.17                            \\ \cline{2-18} 
		\multicolumn{1}{|c|}{}                                        & \bm{$S\_M^T_{RR}$}    & \multicolumn{1}{c|}{1.81}                          & \multicolumn{1}{c|}{0.00}                          & \multicolumn{1}{c|}{3.00}                          & \multicolumn{1}{c|}{1.66}                          & \multicolumn{1}{c|}{\textbf{0.00}}                          & \multicolumn{1}{c|}{1.00}                          & \multicolumn{1}{c|}{0.00}                          & \multicolumn{1}{c|}{0.40}                          & \multicolumn{1}{c|}{2.00}                          & \multicolumn{1}{c|}{1.71}                          & \multicolumn{1}{c|}{0.46}                          & \multicolumn{1}{c|}{1.00}                          & \multicolumn{1}{c|}{1.00}                         & \multicolumn{1}{c|}{0.00}                          & 3.00                          & 17.04                            \\ \cline{2-18} 
		\multicolumn{1}{|c|}{\multirow{-6}{*}{\textbf{Ranking list}}} & \textbf{Average Time (s)} & \multicolumn{1}{c|}{\cellcolor{gray!20}14.33} & \multicolumn{1}{c|}{\cellcolor{gray!20}25.37} & \multicolumn{1}{c|}{\cellcolor{gray!20}18.50} & \multicolumn{1}{c|}{\cellcolor{gray!20}23.21} & \multicolumn{1}{c|}{\cellcolor{gray!20}\textbf{34.92}} & \multicolumn{1}{c|}{\cellcolor{gray!20}21.08} & \multicolumn{1}{c|}{\cellcolor{gray!20}27.38} & \multicolumn{1}{c|}{\cellcolor{gray!20}44.98} & \multicolumn{1}{c|}{\cellcolor{gray!20}13.92} & \multicolumn{1}{c|}{\cellcolor{gray!20}13.11} & \multicolumn{1}{c|}{\cellcolor{gray!20}20.63} & \multicolumn{1}{c|}{\cellcolor{gray!20}16.83} & \multicolumn{1}{c|}{\cellcolor{gray!20}4.69} & \multicolumn{1}{c|}{\cellcolor{gray!20}33.35} & \cellcolor{gray!20}6.33  & \cellcolor{gray!20}21.24    \\ \hline
		\multicolumn{1}{|c|}{}                                        & \bm{$V^T_{equal}$}    & \multicolumn{1}{c|}{0}                             & \multicolumn{1}{c|}{2}                             & \multicolumn{1}{c|}{1}                             & \multicolumn{1}{c|}{1}                             & \multicolumn{1}{c|}{\textbf{1}}                             & \multicolumn{1}{c|}{0}                             & \multicolumn{1}{c|}{2}                             & \multicolumn{1}{c|}{2}                             & \multicolumn{1}{c|}{0}                             & \multicolumn{1}{c|}{2}                             & \multicolumn{1}{c|}{2}                             & \multicolumn{1}{c|}{1}                             & \multicolumn{1}{c|}{3}                            & \multicolumn{1}{c|}{2}                             & 0                             & 19                               \\ \cline{2-18} 
		\multicolumn{1}{|c|}{}                                        & \bm{$S\_M^T_{FMI}$}   & \multicolumn{1}{c|}{0.00}                          & \multicolumn{1}{c|}{1.99}                          & \multicolumn{1}{c|}{1.00}                          & \multicolumn{1}{c|}{1.00}                          & \multicolumn{1}{c|}{\textbf{1.00}}                          & \multicolumn{1}{c|}{0.00}                          & \multicolumn{1}{c|}{2.00}                          & \multicolumn{1}{c|}{1.99}                          & \multicolumn{1}{c|}{0.00}                          & \multicolumn{1}{c|}{2.00}                          & \multicolumn{1}{c|}{1.99}                          & \multicolumn{1}{c|}{0.99}                          & \multicolumn{1}{c|}{3.00}                         & \multicolumn{1}{c|}{2.00}                          & 0.00                          & 18.96                            \\ \cline{2-18} 
		\multicolumn{1}{|c|}{}                                        & \bm{$S\_M^T_{JC}$}    & \multicolumn{1}{c|}{0.00}                          & \multicolumn{1}{c|}{1.99}                          & \multicolumn{1}{c|}{1.00}                          & \multicolumn{1}{c|}{1.00}                          & \multicolumn{1}{c|}{\textbf{1.00}}                          & \multicolumn{1}{c|}{0.00}                          & \multicolumn{1}{c|}{2.00}                          & \multicolumn{1}{c|}{1.99}                          & \multicolumn{1}{c|}{0.00}                          & \multicolumn{1}{c|}{2.00}                          & \multicolumn{1}{c|}{1.98}                          & \multicolumn{1}{c|}{0.99}                          & \multicolumn{1}{c|}{3.00}                         & \multicolumn{1}{c|}{2.00}                          & 0.00                          & 18.95                            \\ \cline{2-18} 
		\multicolumn{1}{|c|}{}                                        & \bm{$S\_M^T_{PR}$}    & \multicolumn{1}{c|}{0.00}                          & \multicolumn{1}{c|}{1.72}                          & \multicolumn{1}{c|}{1.00}                          & \multicolumn{1}{c|}{1.00}                          & \multicolumn{1}{c|}{\textbf{1.00}}                          & \multicolumn{1}{c|}{0.00}                          & \multicolumn{1}{c|}{2.00}                          & \multicolumn{1}{c|}{1.72}                          & \multicolumn{1}{c|}{0.00}                          & \multicolumn{1}{c|}{2.00}                          & \multicolumn{1}{c|}{1.62}                          & \multicolumn{1}{c|}{0.48}                          & \multicolumn{1}{c|}{3.00}                         & \multicolumn{1}{c|}{1.87}                          & 0.00                          & 17.41                            \\ \cline{2-18} 
		\multicolumn{1}{|c|}{}                                        & \bm{$S\_M^T_{RR}$}    & \multicolumn{1}{c|}{0.00}                          & \multicolumn{1}{c|}{1.66}                          & \multicolumn{1}{c|}{1.00}                          & \multicolumn{1}{c|}{1.00}                          & \multicolumn{1}{c|}{\textbf{1.00}}                          & \multicolumn{1}{c|}{0.00}                          & \multicolumn{1}{c|}{2.00}                          & \multicolumn{1}{c|}{1.72}                          & \multicolumn{1}{c|}{0.00}                          & \multicolumn{1}{c|}{2.00}                          & \multicolumn{1}{c|}{1.66}                          & \multicolumn{1}{c|}{0.47}                          & \multicolumn{1}{c|}{3.00}                         & \multicolumn{1}{c|}{1.87}                          & 0.00                          & 17.38                            \\ \cline{2-18} 
		\multicolumn{1}{|c|}{\multirow{-6}{*}{\textbf{Coverage}}}     & \textbf{Average Time (s)} & \multicolumn{1}{c|}{\cellcolor{gray!20}12.00} & \multicolumn{1}{c|}{\cellcolor{gray!20}16.86} & \multicolumn{1}{c|}{\cellcolor{gray!20}38.59} & \multicolumn{1}{c|}{\cellcolor{gray!20}20.17} & \multicolumn{1}{c|}{\cellcolor{gray!20}\textbf{14.26}} & \multicolumn{1}{c|}{\cellcolor{gray!20}46.63} & \multicolumn{1}{c|}{\cellcolor{gray!20}20.33} & \multicolumn{1}{c|}{\cellcolor{gray!20}35.40} & \multicolumn{1}{c|}{\cellcolor{gray!20}34.31} & \multicolumn{1}{c|}{\cellcolor{gray!20}24.72} & \multicolumn{1}{c|}{\cellcolor{gray!20}14.78} & \multicolumn{1}{c|}{\cellcolor{gray!20}21.90} & \multicolumn{1}{c|}{\cellcolor{gray!20}3.33} & \multicolumn{1}{c|}{\cellcolor{gray!20}12.93} & \cellcolor{gray!20}12.71 & \cellcolor{gray!20}21.93    \\ \hline
	\end{tabular}}
\end{table}

The performance of human developers is given in Table~\ref{tab:result_rq4}\footnote{For a faulty version, consistent with the strategy in Section~\ref{subsect:metric}, we analyze the clustering effectiveness only when ``$k$ == $r$'', namely, the number of faults is correctly predicted.}. Take Participant-5 as an example, as shown in the column ``Participant-5'' in Table~\ref{tab:result_rq4}. Participant-5 carries out 9 failure indexing tasks with 3 PMS-based, 3 Ranking list-based, and 3 Coverage-based ones. As for the 3 PMS-based faulty versions, Participant-5 correctly estimate the number of faults on all of them, i.e., the value of the cell (``PMS'', ``$V^T_{equal}$'', ``Participant-5'') is 3. The values of the four clustering metrics, i.e., $S\_M^T_{FMI}$, $S\_M^T_{JC}$, $S\_M^T_{PR}$, $S\_M^T_{RR}$, are both 3.00. Considering that the maximum values of  FMI, JC, PR, and RR are 1, this result indicates that Participant-5 can achieve perfect failure indexing based on PMS. But if based on the other forms of failure representation, Participant-5 performs much worse. Specifically, ``$k$ == $r$'' cannot be obtained on any faulty version based on the representation of ranking lists, and only one faulty version can be properly handled based on the representation of coverage. Moreover, the time cost of manual failure indexing is also different among the three forms of representation. For example, Participant-5 takes 1.17s on average to index each failure when it is represented as PMS images, while takes 34.92s and 14.26s on average when failures are represented as ranking lists and coverage, respectively. That is to say, the proposed failure representation of PMS images enables developers to identify failures with the same/different root cause(s) at a glance, further demonstrating its intuition and \syrevise{friendliness towards human developers}.

The other participants perform similarly to Participant-5: they all tend to achieve better failure indexing with a lower time cost based on the failure representation of PMS, compared with using the other two existing forms of representation. We summarize the human performance based on three forms of failure presentation in the column ``Total'' of Table~\ref{tab:result_rq4}. Totally speaking, regarding the faults number estimation, human developers perform 105.26\% better when using PMS (39 out of 45 times) than when using ranking lists (19 out of 45 times) and coverage (19 out of 45 times). And the four clustering metrics show similar magnitude of improvements. The time costs are 84.65\% and 85.13\% cheaper when using PMS (3.26s) than when using ranking lists (21.24s) and coverage (21.93s). Revisiting the statements at the beginning of this sub-section, we can conclude that the proposed failure representation of PMS has stronger \syrevise{comprehensibility} compared with the two most prevalent and advanced techniques.

At the end of the human study, we interview the participants. They give feedback that based on PMS images, they can immediately justify how many clusters the provided failures should be categorized into, and how they should be divided, because the characteristic of visualization of PMS images is highly in tune with human intuition. But when the other two representations are used, it always takes a lot of time to recognize failures, and the process is very tough.

\section{Discussion}
\label{sect:discussion}

Some interesting topics related to our approach are further discussed in this section.

\subsection{Faulty versions uniquely handled by SURE}
\label{subsect:unique}

\begin{figure}[t]
	\centering
	\includegraphics[width=0.8\linewidth]{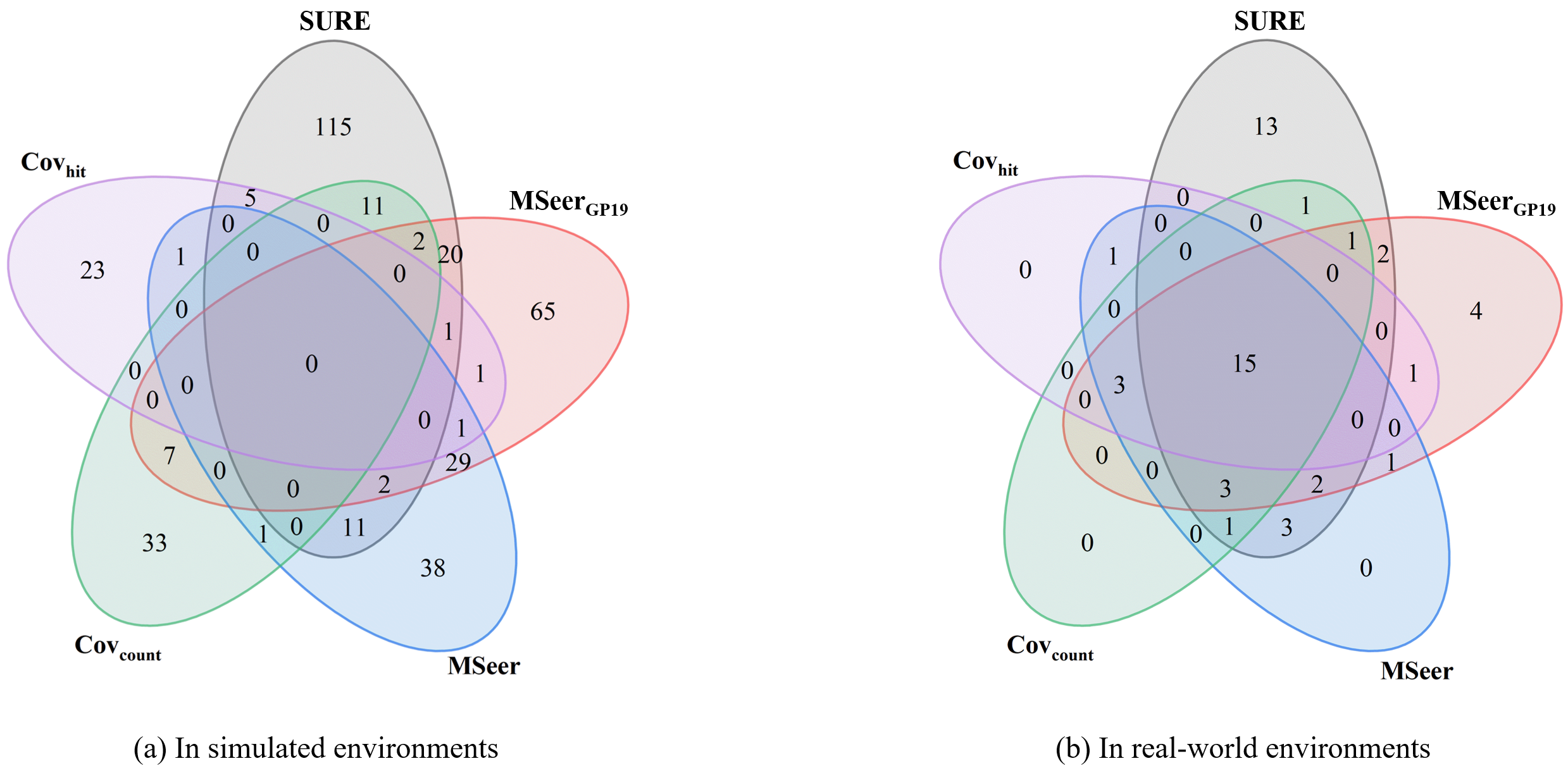}
	\caption{The divergence of the ``$k$ == $r$'' faulty versions}
	\label{fig:discussion_venn}
\end{figure}

In the comparison between SURE and baseline techniques, we find that some faulty versions could be only handled by a specific technique. That is, those ``$k$ == $r$'' faulty versions achieved by different techniques are not exactly the same. It is intuitive that the more faulty versions a failure indexing technique is able to handle solely, the better this technique is. We further inspect SURE and the four baseline techniques from this heuristic aspect, the results are given in Figure~\ref{fig:discussion_venn}.

In Figure~\ref{fig:discussion_venn}(a), we can find that of all SIR faulty versions in the test, 115 can be handled by SURE only, while 65, 38, 33, and 23 can be handled by $MSeer_{GP19}$, $MSeer$, $Cov_{count}$, and $Cov_{hit}$ only, respectively. A similar observation can be drawn from Figure~\ref{fig:discussion_venn}(b): of all Defects4J faulty versions, 13 can be handled by SURE only, while 4, 0, 0, and 0 can be handled by $MSeer_{GP19}$, $MSeer$, $Cov_{count}$, and $Cov_{hit}$ only, respectively. The results reveal that no failure indexing technique can be fully dominated by others, indicating a potential future direction of combining the advantages of different classes of failure proximities together. Despite the respective superiority of various techniques, SURE can solely handle more faulty versions than the other four ones. In particular, it exceeds 76.92\% and 225.00\% by $MSeer_{GP19}$, the best-performed baseline technique, in simulated and real-world environments, respectively.

\subsection{Overhead of SURE}
\label{subsect:overhead}

The time costs of SURE mainly involve three phases, namely, the failure representation, the distance measurement (including the model training and the deployment), and the clustering. According to our analyses, as for the failure representation, SURE takes 3.99 minutes and 5.90 minutes on average to collect run-time memory information during the execution of a failed test case, and then takes 0.18s and 0.27s on average to convert a set of memory information to a PMS image, on SIR and Defects4J faulty versions, respectively. As for the distance measurement, SURE takes 23 minutes and 35 seconds to train the Siamese-based model. Once the model is trained, 0.01s and 0.06s on average are taken to predict the distance between a pair of PMS images on SIR and Defects4J faulty versions, respectively. After these two steps are ready, the clustering process typically takes only a few seconds. To summarize, the overhead of SURE mainly lies in querying run-time memory information at preset breakpoints.

As pioneers have pointed out, failure indexing is essential yet very costly when it comes to manual debugging in the real world. ``\emph{Experienced developers can manually examine every failure and determine the culprit fault, but this is apparently too expensive}'', pioneers concluded~\cite{liu2008systematic}. In contrast, SURE can complete this task automatically and hence can save a great deal of manual effort, which is far less expensive than manual jobs. In particular, SURE represents a failed test case as a PMS image, which can be smoothly \syrevise{comprehended} by human developers at a glance and thus can boost the \syrevise{comprehensibility} of failure indexing outcomes. This can be of great importance because if developers are not convinced by failure indexing outcomes, they will take much time to verify their correctness. Even though coverage vectors (the failure representation of the CC-based strategy) and statement ranking lists (the failure representation of the SD-based strategy) can also provide insights for developers to \syrevise{comprehend} failure indexing outcomes to an extent, they can be human labor-intensive. To put it another way, we transfer the costs that would have been borne by humans to machines. It is true that as compared with CC and SD-based techniques, SURE needs higher costs. However, in return, SURE delivers better performance and convenience. Therefore, the cost of SURE is acceptable: more sophisticated fingerprinting is naturally accompanied by higher overhead, i.e., \textbf{``\emph{no free lunch}''}~\cite{liu2008systematic}.

\section{Threats to Validity}
\label{sect:threats}

Our experiments are subject to several threats to validity.

Threats to internal validity relate to the faulty versions we focus on to determine the effectiveness of a failure indexing technique. In the experimental evaluation, we only consider those faulty versions that satisfy the condition of ``$k$ == $r$'', i.e., where the predicted number of faults $k$ is equal to the real number of faults $r$. Actually, ``$k$ == $r$'' is ideal but not necessary for parallel debugging to work in practice. This is because if $k$ exceeds $r$, $k$ developers will be employed to locate $r$ faults, and parallel debugging can work at the cost of human labor ($k$ - $r$ developers are redundant). On the contrary, if $k$ is less than $r$, parallel debugging can also work at the cost of more than one iteration of debugging. Filtering these faulty versions is because comparing $k$ generated clusters with $r$ oracle clusters when ``$k$ != $r$'' is difficult, and moreover, prior studies have pointed out that the performance of a failure indexing technique can be mainly determined by those ``$k$ == $r$'' faulty versions~\cite{gao2019mseer,song2022comprehensive}, that is, the contribution of ``$k$ != $r$'' ones is marginal. Therefore, the threat is acceptable. Nonetheless, we plan to further take this imperfect situation into account in our future work, for more comprehensive evaluation of failure indexing techniques.

Threats to external validity relate to the generalization capability of SURE. Specifically, given that the evaluation of the effectiveness of SURE is carried out empirically, our results may not be extended to all programs. In the experiments, we use nine projects from two open-source platforms, which are written in different languages (C and Java) and with various functionalities, thus mitigating the threat to an extent. In particular, we use only 30\% of the simulated faults to train the model, and use the remaining 70\% of the simulated as well as real-world faults to test. This allows us to have higher confidence with respect to the applicability of SURE.

Threats to construct validity relate to the adopted evaluation metrics.  We use external metrics (i.e., FMI, JC, PR, and RR) to measure the clustering effectiveness. Despite the fact that these metrics have been extensively used by previous works, we consider using more diverse metrics in our future work for further evaluation.

\section{Related Work}
\label{sect:relatedwork}

\iffalse
Failure indexing has been a longstanding task in the field of software testing and debugging. There are generally two classes of failure indexing according to the form of failures: bug report-oriented and fail test case-oriented. This paper is under the scenario of dynamic testing, i.e., failures are in the form of failed test cases. Therefore, in this section, we introduce related works with regard to fail test case-oriented failure indexing.
\fi

It is well-recognized that the failure proximity essentially underpins failure indexing techniques. Liu et al. systematically summarized the status of the community of failure proximities, pointing out that CC-based and SD-based strategies can deliver good results~\cite{liu2008systematic}. In fact, these two are mainstream tactics in failure indexing to date, and numerous studies have emerged around them in the last two decades.

Let us first focus on the CC-based strategy. As a very early work, Podgurski et al. suggested using execution profiles, including code coverage, as the signature of failures~\cite{podgurski2003automated}. Since then, this tactic has become popular gradually. For example, Liu et al. formally presented the definition of the CC-based failure proximity, and demonstrated its capability by experiments~\cite{liu2008systematic}. Högerle et al. constructed failed test coverage matrices, and used the Weil-Kettler algorithm to rearrange the mentioned coverage matrix into a block diagonal matrix thus achieving failure clustering~\cite{hogerle2014more}. Also based on failed test coverage matrices, Steimann and Frenkel utilized two partitioning procedures from integer linear programming to cluster failures, for breaking down the fault localization problem into smaller ones~\cite{steimann2012improving}. Huang et al. were concerned with the impact of failure indexing on the effectiveness of multi-fault localization. They conducted an empirical study to explore this topic on the basis of taking code coverage as the representation of failures~\cite{huang2013empirical}. Wu et al. also employed code coverage as the signature of failures to perform failure clustering. They iteratively chose the cluster of failures with the highest density to conduct single-fault localization, until all faults were fixed~\cite{wu2020fatoc}.

Later, the SD-based strategy, a more sophisticated solution of failure proximity, was proposed by Liu and Han~\cite{liu2006failure}, which regards two failures as similar if they suggest roughly the same fault location. Specifically, they used SOBER, a statistical model-based fault localization technique at the predicate granularity~\cite{liu2005sober}, to complete the fault location suggestion process. From then on, the SD-based strategy attracted more and more attention from academia. For example, Jones et al. used Tarantula~\cite{jones2005empirical}, an SBFL technique working at the statement granularity, to suggest finer-grained fault location. As a result, the suspiciousness ranking list that reflects the possibility of each program statement being faulty is utilized to represent failures~\cite{jones2007debugging}. Following the workflow of the SD-based strategy, Gao and Wong proposed MSeer, which employed Crosstab~\cite{wong2012towards}, an empirically promising SBFL technique, to produce the suspiciousness ranking list that represents failures~\cite{gao2019mseer}. MSeer is also the state-of-the-art technique to date in the field of failure indexing. Song et al. investigated the impact of the adopted SBFL techniques on the effectiveness of SD-based failure indexing. Their conclusion demonstrated that MSeer can be further enhanced by replacing Crosstab with other SBFL formulas, such as the one evolved by genetic programming, GP19~\cite{song2022comprehensive}.

Despite the integration of fault localization techniques, the resource on which the SD-based strategy relies is still only coverage. If failures that are triggered by distinct faults have the same coverage, neither the CC-based nor the SD-based strategy can work well. However, this situation has been demonstrated to be very common in practice, which causes performance degradation in CC and SD strategies. Moreover, the failure representation (code coverage vectors for CC and suspiciousness ranking lists for SD) is also tough and time-consuming to \syrevise{comprehend} by human developers, which hinders human comprehension of failure indexing results. In this paper, we propose SURE, a novel failure indexing technique, to tackle the aforementioned threats to existing approaches. With the support of run-time memory information, SURE remarkably improves the effectiveness of failure indexing, and PMS images, the novel form of failure representation adopted by SURE, help human developers comprehend the result of failure indexing better.

There are some recent works that introduce external profiles to support failure indexing, such as code-independent features in regression testing~\cite{golagha2019failure}, as well as code features and historical features in continuous integration~\cite{an2022automatically}. We do not consider such types of studies since they go beyond our research scope: 1) their source information cannot be always available, and 2) This paper focuses on failure indexing in the context of multi-fault debugging.

\section{Conclusion}
\label{sect:conclusion}

In this paper, we propose the program memory-based failure proximity, and based on that propose a novel failure indexing technique, SURE. Experimental results demonstrate the high competitiveness of SURE: it can achieve 101.20\% and 41.38\% improvements in faults number estimation, as well as 105.20\% and 35.53\% improvements in clustering, compared with the state-of-the-art technique in this field to date, in simulated and real-world environments, respectively. SURE also provides human developers with insights to better comprehend the failure indexing results. Our human study involving 15 participants shows that the form of failure representation of SURE, i.e., PMS images, can boost human developers' \syrevise{comprehension} of failure indexing results by 105.26\% compared with the two most prevalent and advanced strategies, while the time cost is reduced by 84.65\% and 85.13\%, respectively.

In the future, we plan to dig deeper into the characteristics of various forms of failure proximities, and try to combine their advantages for better extracting the signature of failures. A more extensive experiment with a larger scale of benchmarks as well as more diverse evaluation metrics is also being considered.

\iffalse
\section*{Acknowledgments}

This work was partially supported by the National Natural Science Foundation of China under the grant numbers \synote{PLACEHOLDER} and \synote{PLACEHOLDER} . And the numerical calculations in this work have been partially done on the supercomputing system in the Supercomputing Center of Wuhan University.
\fi

\bibliographystyle{ACM-Reference-Format}
\bibliography{sample-base}

\appendix

\end{document}